\documentclass[12pt]{elsarticle}

\usepackage{amssymb}

\usepackage{graphicx}
\usepackage{amsmath,amssymb}
\usepackage{mathrsfs}
\usepackage{subfigure}
\usepackage{color}
\usepackage{multirow}
\usepackage{bbold}
\biboptions{sort&compress}

\journal{Journal of magnetism and magnetic materials}

\begin{document}

\begin{frontmatter}



\title{Collinear antiferromagnetic phases of a frustrated spin-$\frac{1}{2}$ $J_{1}$--$J_{2}$--$J_{1}^{\perp}$ Heisenberg model on an $AA$-stacked bilayer honeycomb lattice}


\author{P H Y Li$^{1,2}$ and R F Bishop$^{1,2}$}
\ead{peggyhyli@gmail.com; raymond.bishop@manchester.ac.uk}

\address{$^{1}$ School of Physics and Astronomy, Schuster Building, The University of Manchester, Manchester, M13 9PL, UK}

\address{$^{2}$ School of Physics and Astronomy, University of Minnesota, 116 Church Street SE, Minneapolis, Minnesota 55455, USA}

\begin{abstract}
The regions of stability of two collinear quasiclassical phases within the zero-temperature quantum phase diagram of the spin-$\frac{1}{2}$ $J_{1}$--$J_{2}$--$J_{1}^{\perp}$ model on an $AA$-stacked bilayer honeycomb lattice  are investigated using the coupled cluster method (CCM).  The model comprises two monolayers in each of which the spins, residing on honeycomb-lattice sites, interact via both nearest-neighbor (NN) and frustrating next-nearest-neighbor isotropic antiferromagnetic (AFM) Heisenberg exchange iteractions, with respective strengths $J_{1} > 0$ and $J_{2} \equiv \kappa J_{1}>0$.  The two layers are coupled via a comparable Heisenberg exchange interaction between NN interlayer pairs, with a strength $J_{1}^{\perp} \equiv \delta J_{1}$.  The complete phase boundaries of two quasiclassical collinear AFM phases, namely the N\'{e}el and N\'{e}el-II phases on each monolayer, with the two layers coupled so that NN spins between them are antiparallel, are calculated in the $\kappa \delta$ half-plane with $\kappa > 0$.  Whereas on each monolayer in the N\'{e}el state all NN pairs of spins are antiparallel, in the N\'{e}el-II state NN pairs of spins on zigzag chains along one of the three equivalent honeycomb-lattice directions are antiparallel, while NN interchain spins are parallel.  We calculate directly in the thermodynamic (infinite-lattice) limit both the magnetic order parameter $M$ and the excitation energy $\Delta$ from the $s^{z}_{T}=0$ ground state to the lowest-lying $|s^{z}_{T}|=1$ excited state (where $s^{z}_{T}$ is the total $z$ component of spin for the system as a whole, and where the collinear ordering lies along the $z$ direction), for both quasiclassical states used (separately) as the CCM model state, on top of which the multispin quantum correlations are then calculated to high orders ($n \leq 10$) in a systematic series of approximations involving $n$-spin clusters.  The sole approximation made is then to extrapolate the sequences of $n$th-order results for $M$ and $\Delta$ to the exact limit, $n \to \infty$.
\end{abstract}


\end{frontmatter}


\section{Introduction}
\label{introd_sec}
Of all the bipartite lattices in two dimensions, and in which all of the sites and all of the edges are equivalent to one another, the honeycomb lattice has the lowest value of the coordination number.  Hence, it is expected to show the greatest effect of quantum fluctuations when populated by spins interacting via antiferromagnetic (AFM), isotropic Heisenberg interactions between nearest-neighbor (NN) pairs only.  We also expect the largest deviations from classical behavior to occur when the spin quantum number $s$ takes the lowest value, $s=\frac{1}{2}$.  It is only natural therefore for spin-$\frac{1}{2}$ models on the (infinite) two-dimensional (2D) honeycomb lattice to play a special role in the study of quantum phase transitions (QPTs), since lower values of the system dimensionality $d$ also tend to favor the enhancement of quantum effects.  One of the most directly observable such effects of quantum fluctuations will be to reduce the value of the order parameter $M$ (defined to be the average local onsite magnetization or, equivalently, the sublattice magnetization for bipartite lattices) from its classical value equal to $s$, either to zero or to some nonzero value.

For the simplest, unfrustrated Heisenberg antiferromagnet (HAF) with NN exchange interactions only, all of equal strength $J_{1} > 0$, on a 2D monolayer honeycomb lattice, it is by now well established that the perfect N\'{e}el long-range order (LRO) that exists in the classical ($s \to \infty$) limit, i.e., $M=s$, is not destroyed totally by quantum fluctuations for any finite value of $s$.  Rather, $M$ is reduced from its classical value, but to a value still greater than zero.  For example, for the extreme quantum case $s=\frac{1}{2}$, the N\'{e}el order parameter is reduced by nearly half from its classical value, taking the value $M \approx 0.27$ \cite{Richter:2004_triang_ED,DJJFarnell:2014_archimedeanLatt}.  Accordingly, it is now interesting to enquire about how the N\'{e}el LRO on the honeycomb monolayer might be destroyed by the inclusion of additional competing interactions to the NN AFM bonds.  Two straightforward means to do so are now considered, one involving frustration and other without.

The first method is to include isotropic AFM Heisenberg exchange bonds between next-nearest-neighbor (NNN) pairs of spins, all with equal strength, $J_{2} \equiv \kappa J_{1}$.  If $J_{2}>0$, the $J_{2}$ bonds clearly act to frustrate the N\'{e}el order promoted by the $J_{1}>0$ bonds.  The resulting $J_{1}$--$J_{2}$ model thus obtained has been much studied on the honeycomb lattice, particularly for the case $s=\frac{1}{2}$, using a large cross-section of available analytical and numerical techniques \cite{Rastelli:1979_honey,Mattsson:1994_honey,Fouet:2001_honey,Mulder:2010_honey,Ganesh:2011_honey,Ganesh:2011_honey_errata,Clark:2011_honey,Reuther:2011_honey,Albuquerque:2011_honey,Mosadeq:2011_honey,Oitmaa:2011_honey,Mezzacapo:2012_honey,Li:2012_honey_full,Bishop:2012_honeyJ1-J2,RFB:2013_hcomb_SDVBC,Zhang:2013_honey,Ganesh:2013_honey_J1J2mod-XXX,Zhu:2013_honey_J1J2mod-XXZ,Gong:2013_J1J2mod-XXX,Yu:2014_honey_J1J2mod}.  The second method to include additional competing bonds, now without frustration, is to take two identical honeycomb monolayers and arrange them into an $AA$-stacked bilayer (i.e., with each site of one monolayer placed immediately above its equivalent on the other monolayer), and now add a NN interlayer Heisenberg exchange coupling so that all such bonds have equal strength, $J_{1}^{\perp} \equiv \delta J_{1}$.  Such interlayer $J_{1}^{\perp}$ bonds do not directly frustrate the N\'{e}el LRO promoted by the intralayer $J_{1}$ bonds.  Indeed, at the classical level ($s \to \infty$) they have no effect at all.  Nevertheless, for finite values of $s$, the $J_{1}$ and $J_{1}^{\perp}$ bonds are in competition with one another since the $J_{1}^{\perp}$ bonds acting alone will promote the formation of interlayer NN dimers (i.e., spin-singlet pairs in the case $J_{1}^{\perp}>0$, and spin-triplet pairs when $J_{1}^{\perp}<0$).  Thus, the inclusion of the $J_{1}^{\perp}$ bonds leads to a competition between a phase with N\'{e}el magnetic LRO on each monolayer and a nonclassical paramagnetic phase of the valence-bond crystalline (VBC) kind and formed of interlayer dimers.  The resulting $J_{1}$--$J_{1}^{\perp}$ model, and its N\'{e}el to dimer quantum phase transition, has been studied on the bilayer honeycomb lattice, for the cases of spins with $s=\frac{1}{2},1,\frac{3}{2}$, using both exact stochastic series expansion quantum Monte Carlo (QMC) simulation algorithms and an approximate analysis based on a bond operator method that has been generalized to arbitrary values of $s$ \cite{Ganesh:2011_honey_bilayer_PRB84}.

More recently there have been some initial studies of the so-called $J_{1}$--$J_{2}$--$J_{1}^{\perp}$ model on the bilayer honeycomb lattice, in which both types of competition discussed above to destroy N\'{e}el order act simultaneously \cite{Zhang:2014_honey_bilayer,Arlego:2014_honey_bilayer,Bishop:2017_honeycomb_bilayer_J1J2J1perp,Li:2017_honeycomb_bilayer_J1J2J1perp_Ext-M-Field}.  In the earlier work \cite{Zhang:2014_honey_bilayer,Arlego:2014_honey_bilayer} the model was studied using Schwinger-boson mean field theory, augmented by the exact diagonalization of a relatively small (24-site) cluster, linear spin-wave theory, and a calculation of the spin-triplet energy gap using a dimer series expansion carried out to relatively low (viz., fourth) orders.  In our own later work \cite{Bishop:2017_honeycomb_bilayer_J1J2J1perp,Li:2017_honeycomb_bilayer_J1J2J1perp_Ext-M-Field} we studied the model for the case $s=\frac{1}{2}$ using a high-order implementation of a fully microscopic quantum many-body theory technique, namely the coupled cluster method (CCM), which yielded accurate results for the ground-state (GS) energy per spin, the N\'{e}el magnetic order parameter $M$, the excitation energy $\Delta$ from the $s_{T}^{z}=0$ ground state to the lowest-lying $|s_{T}^{z}|=1$ excited state (where $s_{T}^{z}$ is the total $z$ component of spin for the system as a whole, and where the collinear ordering lies along the $z$ direction), and the zero-field transverse (uniform) magnetic susceptibility $\chi$ in the N\'{e}el phase.  We thus obtained in particular an accurate estimate for the full phase boundary of the N\'{e}el phase in the quadrant with $\kappa > 0$ and $\delta > 0$ of the $\kappa \delta$ plane of the zero-temperature ($T=0$) quantum phase diagram.

We note that the CCM \cite{Coester:1958_ccm,Coester:1960_ccm,Cizek:1966_ccm,Kummel:1978_ccm,Bishop:1978_ccm,Bishop:1982_ccm,Arponen:1983_ccm,Bishop:1987_ccm,Arponen:1987_ccm,Arponen:1987_ccm_2,Bartlett:1989_ccm,Arponen:1991_ccm,Bishop:1991_TheorChimActa_QMBT,Bishop:1998_QMBT_coll,Zeng:1998_SqLatt_TrianLatt,Fa:2004_QM-coll} has itself been applied with great success to a very wide array of systems in quantum magnetism, in almost all of which it has yielded results which are either the most accurate or among the most accurate available.  In particular, these encompass a considerable number of applications to a variety of frustrated monolayer honeycomb-lattice models \cite{DJJF:2011_honeycomb,PHYLi:2012_honeycomb_J1neg,Bishop:2012_honey_circle-phase,Li:2012_honey_full,Bishop:2012_honeyJ1-J2,RFB:2013_hcomb_SDVBC,Bishop:2014_honey_XY,Li:2014_honey_XXZ,Bishop:2014_honey_XXZ_nmp14,Bishop:2015_honey_low-E-param,Bishop:2016_honey_grtSpins,Li:2016_honeyJ1-J2_s1,Li:2016_honey_grtSpins}, including the spin-$\frac{1}{2}$ $J_{1}$--$J_{2}$ model itself \cite{Bishop:2012_honeyJ1-J2,RFB:2013_hcomb_SDVBC}.  Apart from the N\'{e}el AFM state exhibited by this model at low values of the frustration parameter $\kappa \equiv J_{2}/J_{1}$, several accurate calculations (and see, e.g., Refs.\ \cite{Fouet:2001_honey,Albuquerque:2011_honey,Mezzacapo:2012_honey,Yu:2014_honey_J1J2mod}), including those using the CCM \cite{Li:2012_honey_full,Bishop:2012_honeyJ1-J2,RFB:2013_hcomb_SDVBC}, show that it also exhibits another quasiclassical phase with collinear magnetic LRO, viz., the so-called N\'{e}el-II phase described in more detail in Sec.\ \ref{model_sec}, for higher values of $\kappa$.  In between these two quasiclassical magnetic phases there is also broad agreement between calculations based on a variety of different techniques that the system is paramagnetic, with VBC order of the plaquette (PVBC) and/or staggered dimer (SDVBC or, equivalently, lattice nematic) type.  There have also been hints of possible small regions of $\kappa$ beyond the N\'{e}el regime where the stable GS phase may be a quantum spin liquid (QSL).

In view of the richness and complexity of the spin-$\frac{1}{2}$ $J_{1}$--$J_{2}$ model on a monolayer honeycomb lattice, it is clearly of great interest to investigate the comparable spin-$\frac{1}{2}$ $J_{1}$--$J_{2}$--$J_{1}^{\perp}$ model on the honeycomb bilayer.  Of particular interest will be to investigate the stability and sensitivity of the various GS phases exhibited by the monolayer to the degree of interlayer coupling, $\delta \equiv J_{1}^{\perp}/J_{1}$, that is present.  To date the only calculations performed on this model (in $AA$ stacking), to our knowledge, have been to investigate the stability of the N\'{e}el phase in the $\kappa \delta$ plane for $\kappa > 0$ and $\delta > 0$ \cite{Zhang:2014_honey_bilayer,Arlego:2014_honey_bilayer,Bishop:2017_honeycomb_bilayer_J1J2J1perp,Li:2017_honeycomb_bilayer_J1J2J1perp_Ext-M-Field}.  In the present paper our aim is to extend those earlier preliminary calculations to include both quasiclassical AFM GS phases (viz., the N\'{e}el and N\'{e}el-II phases) present when $\delta = 0$, and to investigate their realms of stability in the entire $\kappa \delta$ half-plane with intralayer frustration (i.e., $\kappa > 0$).  In so doing we will shed considerable light on the extraordinary sensitivity of {\it both} phases on the honeycomb monolayer, specifically by showing explicitly how their corresponding phase boundaries change rapidly as functions of $\kappa$ in the region of small interlayer coupling $\delta$.

The plan of the rest of the paper is as follows.  We first describe the $J_{1}$--$J_{2}$--$J_{1}^{\perp}$ model itself in Sec.\ \ref{model_sec}, including a description of its main features in the limiting case, $J_{1}^{\perp} \to 0$, of the monolayer.  In Sec.\ \ref{ccm_sec} we briefly review the main features of the CCM, before presenting our results for the N\'{e}el and N\'{e}el-II phases obtained from using it in Sec.\ \ref{results_sec}.  The results are then discussed and summarized in Sec.\ \ref{discuss_summary_sec}.

\section{The model}
\label{model_sec}
The $J_{1}$--$J_{2}$--$J_{1}^{\perp}$ model on the
bilayer honeycomb lattice is specified by the Hamiltonian
\begin{equation}
\begin{aligned}
H & =  J_{1}\sum_{{\langle i,j \rangle},\alpha} \mathbf{s}_{i,\alpha}\cdot\mathbf{s}_{j,\alpha} + 
J_{2}\sum_{{\langle\langle i,k \rangle\rangle},\alpha} \mathbf{s}_{i,\alpha}\cdot\mathbf{s}_{k,\alpha}  \\
& \quad + J_{1}^{\perp}\sum_{i} \mathbf{s}_{i,A}\cdot\mathbf{s}_{i,B} \\
& \equiv  J_{1}h(\kappa,\delta)\,; \quad  \kappa \equiv J_{2}/J_{1}\,, \quad \delta \equiv J_{1}^{\perp}/J_{1}\,,
\label{H_eq}
\end{aligned}
\end{equation}
where the index $i$ labels the sites on each (horizontal) monolayer
(i.e., in $AA$ stacking, so that corresponding sites $i$ on the top
layer lie vertically above those on the lower layer), and the index
$\alpha = A,B$ labels the two layers.  Each site ($i, \alpha$) carries
a spin-$s$ particle described by the usual SU(2) spin operators
${\bf
  s}_{i,\alpha}\equiv(s^{x}_{i,\alpha},s^{y}_{i,\alpha},s^{z}_{i,\alpha})$,
with ${\bf s}^{2}_{i,\alpha} = s(s+1)\mathbb{1}$.  We restrict ourselves here
to the extreme quantum case, $s=\frac{1}{2}$.  The first two sums in
Eq.\ (\ref{H_eq}) over $\langle i,j \rangle$ and
$\langle \langle i,k \rangle \rangle$ run respectively over all NN and
NNN intralayer pairs of spins on each monolayer honeycomb lattice,
counting each Heisenberg bond (with strengths $J_{1}$ and $J_{2}$
respectively) once and once only.  The third sum in Eq.\ (\ref{H_eq})
describes the interlayer Heisenberg bonds, with strength
$J_{1}^{\perp}$, between NN pairs of spins across the two $AA$-stacked
monolayers (i.e., at the same horizontal site index $i$).
\begin{figure*}[t]
\mbox{
\subfigure[]{\includegraphics[width=2.8cm]{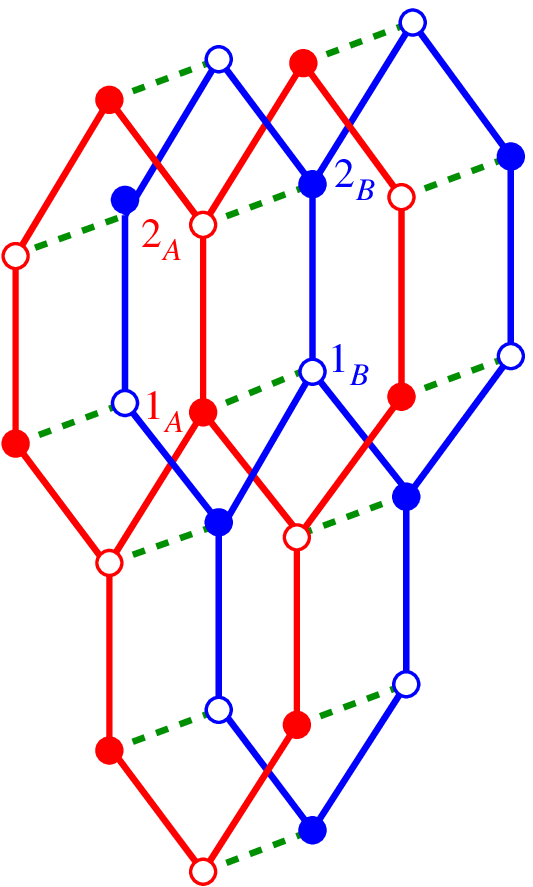}}
\hspace{0.1cm}
\subfigure[]{\includegraphics[width=3.7cm]{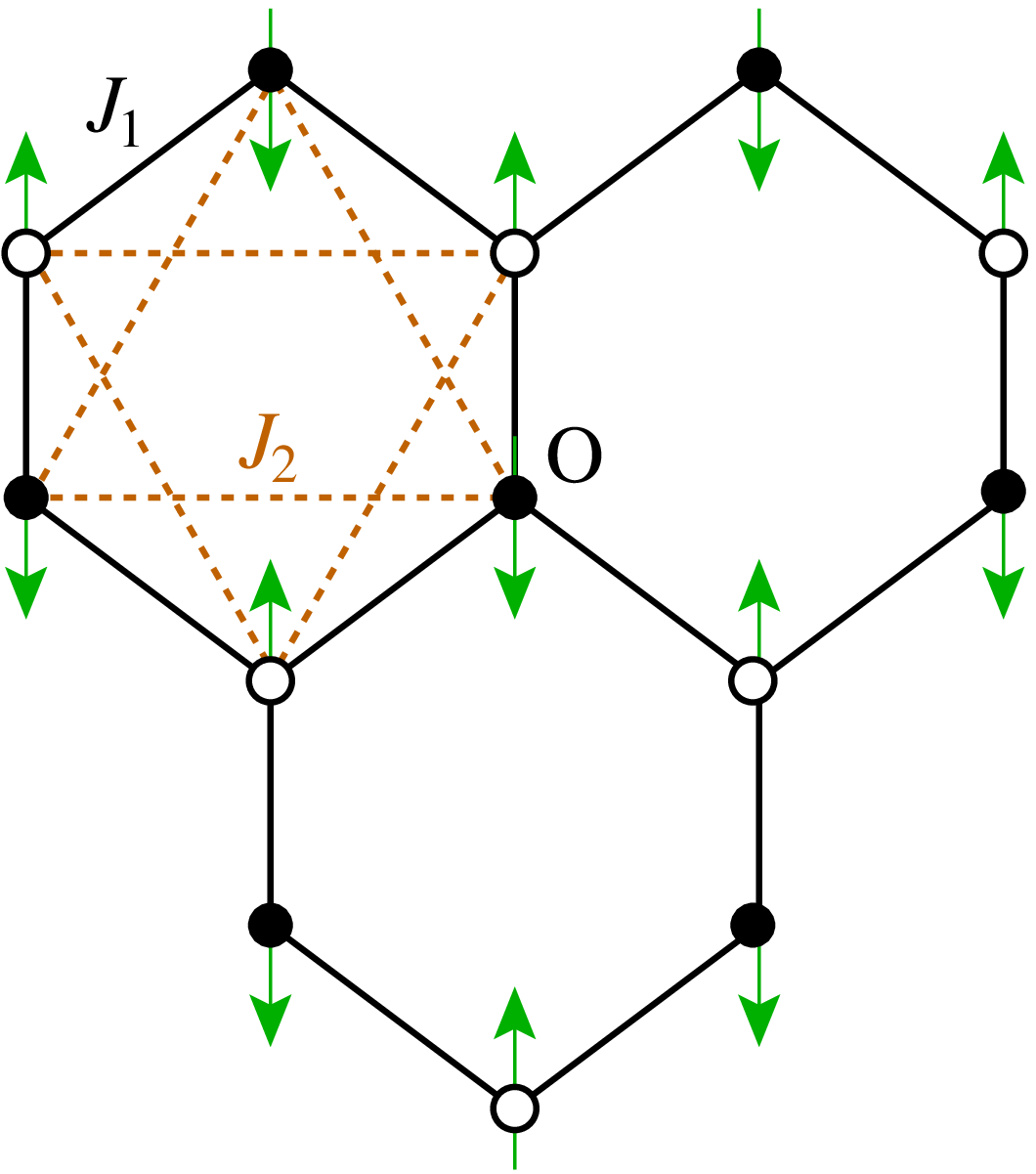}}
\hspace{0.1cm}
\subfigure[]{\includegraphics[width=3.7cm]{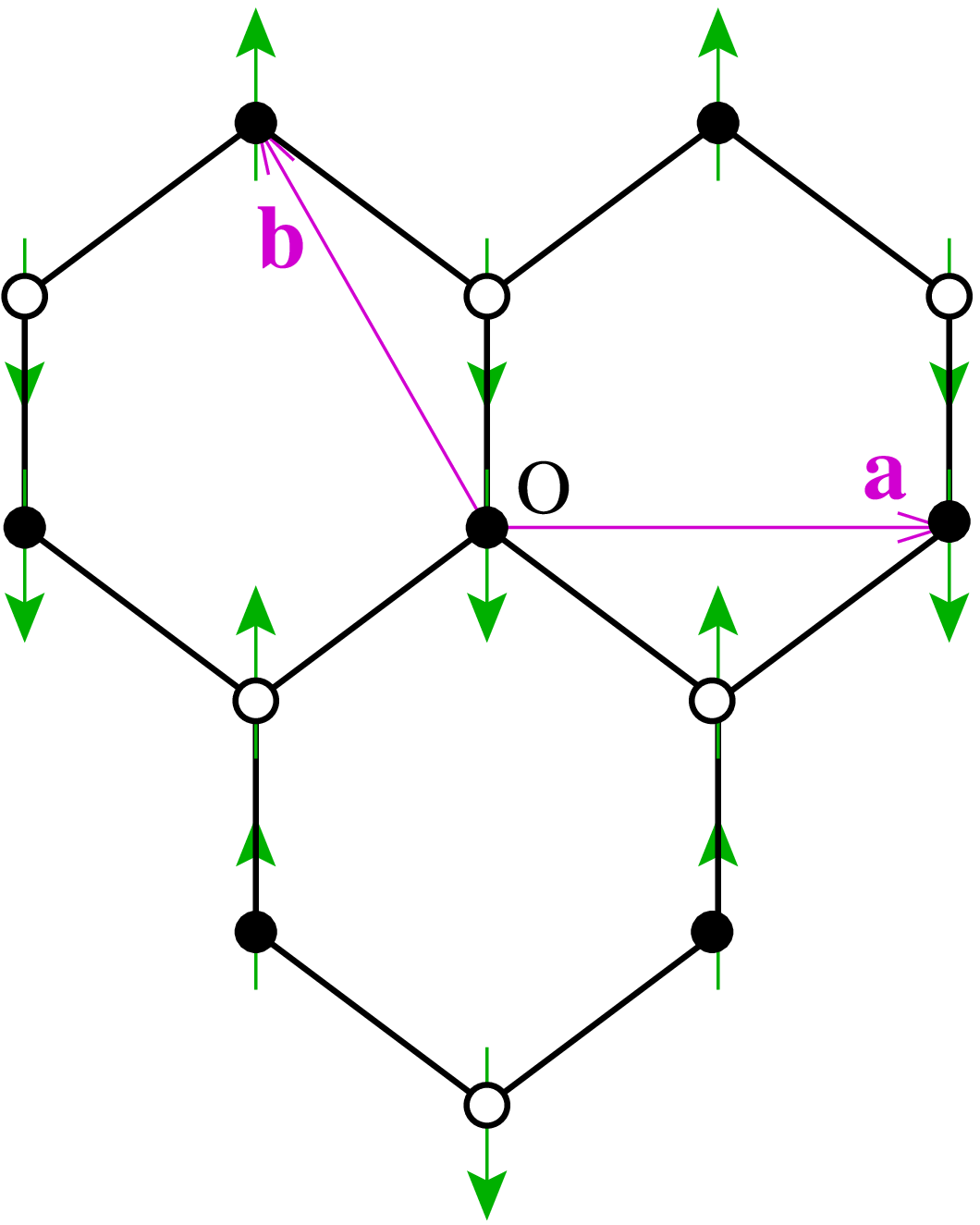}}
\hspace{0.1cm}
\subfigure{\includegraphics[width=2.5cm]{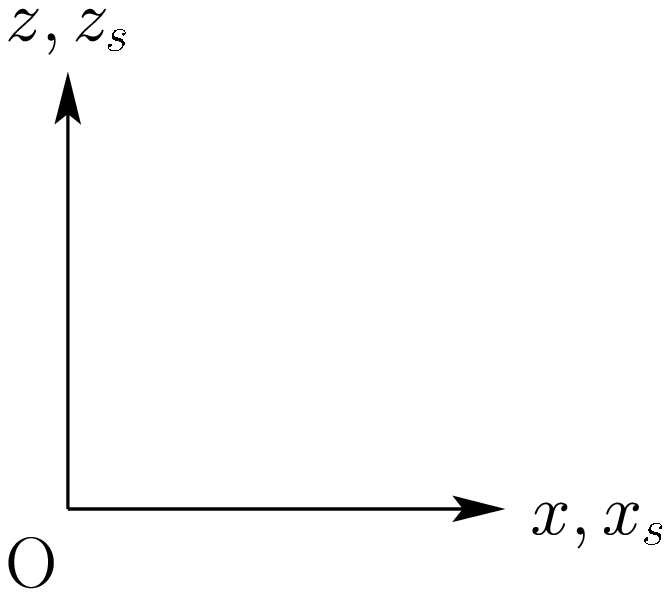}}
}
  \caption{The $J_{1}$--$J_{2}$--$J_{1}^{\perp}$ model on the bilayer honeycomb 
    lattice, showing (a) the two layers $A$ (red) and $B$ (blue), the nearest-neighbor (NN) bonds ($J_{1} = $ -----; $J_{1}^{\perp} = $ - - -)  and the four sites ($1_{A}, 2_{A}, 1_{B}, 2_{B}$) of the unit cell; (b) the intralayer bonds ($J_{1} = $ -----; $J_{2} = $ - - -) on each monolayer and the monolayer N\'{e}el state; and (c) the triangular Bravais lattice vectors ${\bf a}$ and ${\bf b}$, and one of the three equivalent monolayer N\'{e}el-II states.  Sites ($1_{A}, 2_{B}$) and ($2_{A}, 1_{B}$) on the two triangular lattices of each monolayer are shown by filled and empty circles respectively, and the spins are represented by the (green) arrows on the lattice sites.  For both states on the bilayer, spins on NN sites between the two layers are antiparallel.}
\label{model_pattern}
\end{figure*}

We are interested here in the case when both intralayer bonds are AFM in nature (i.e., $J_{1}>0$, $J_{2}>0$).  Since the parameter $J_{1}$ simply sets the overall energy scale, we may write the Hamiltonian as in the last line in Eq.\ (\ref{H_eq}), where the relevant parameters are thus $\kappa \equiv J_{2}/J_{1}$ and $\delta \equiv J_{1}^{\perp}/J_{1}$.  The lattice and the Heisenberg exchange bonds are illustrated in Figs.\ \ref{model_pattern}(a) and \ref{model_pattern}(b).
We thus restrict attention here to investigating the stability of the
two quasiclassical collinear AFM phases in the $\kappa \delta$
half-plane with $\kappa > 0$.  The spin patterns of these two phases
on each monolayer, viz., the N\'{e}el and N\'{e}el-II states, are
shown respectively in Figs.\ \ref{model_pattern}(b) and
\ref{model_pattern}(c).  Whereas the N\'{e}el state has all three NN
spins to a given spin antiparallel to it, the N\'{e}el-II state has
two such spins antiparallel and one parallel.  The N\'{e}el state may
equivalently be described as having AFM sawtooth (or zigzag) chains
along all three of the equivalent honeycomb-lattice directions.  By
contrast, each N\'{e}el-II state has AFM sawtooth along only one of
the honeycomb directions [e.g., along the $\hat{x}$ direction in Fig.\
\ref{model_pattern}(c)], and now with NN spins on adjacent chains
parallel to each other.  There are thus two other equivalent
N\'{e}el-II states to that shown in Fig.\ \ref{model_pattern}(c),
obtained from it by rotations in the $xz$ plane by $\pm 120^{\circ}$
about the center of any hexagon.  The N\'{e}el-II states thus break
the lattice rotational symmetry that is preserved by the N\'{e}el
state by contrast.  Whereas the N\'{e}el state has a 2-site unit cell
structure, the N\'{e}el-II state has a 4-site structure.  Since the
N\'{e}el-II state also comprises collinear stripes of parallel spins
[i.e., along lines parallel to the $z$-axis in Fig.\
\ref{model_pattern}(c)] that alternate in direction, it is also
sometimes called the collinear striped AFM phase in the literature.
We note that we prefer to avoid this terminology, since it is open to
considerable confusion with other AFM states on the honeycomb lattice
that are also known as striped states (and see, e.g., Ref.\
\cite{Li:2012_honey_full}) and which comprise sawtooth chains of
parallel spins that alternate in direction.

While the honeycomb lattice is bipartite, it is non-Bravais.  It
comprises two interpenetrating triangular Bravais sublattices 1 and
2, with lattice vectors ${\mathbf a}=\sqrt{3}d\hat{x}$ and
${\mathbf b}=\frac{1}{2}d(-\sqrt{3}\hat{x}+3\hat{z})$, as shown in
Fig.\ \ref{model_pattern}(c), where each monolayer is defined to lie
in an $xz$ plane, as illustrated in Fig.\ \ref{model_pattern}, and $d$
is the honeycomb lattice spacing (i.e., the separation distance
between NN pairs on the hexagonal lattice).  Each monolayer unit cell
$l$ at position vector
${\mathbf R}_{l}=m_{l}{\mathbf a}+n_{l}{\mathbf b}$, where
$m_{l},n_{l} \in \mathbb{Z}$, thus comprises the two sites at
${\mathbf R}_{l}$ on sublattice 1 and ${\mathbf R}_{l}+d\hat{z}$ on
sublattice 2.  The corresponding four sites of the $AA$-stacked
bilayer honeycomb lattice unit cell are shown in Fig.\
\ref{model_pattern}(a).

Clearly, the reciprocal lattice vectors that correspond to the
real-space vectors ${\mathbf a}$ and ${\mathbf b}$ may be taken as
$\boldsymbol{\alpha}=2\pi(\sqrt{3}\hat{x}+\hat{z})/(3d)$ and
$\boldsymbol{\beta}=4\pi/(3d)\hat{z}$.  The parallelograms formed by
the pairs of vectors (${\mathbf a},{\mathbf b}$) and
($\boldsymbol{\alpha},\boldsymbol{\beta}$) are thus the Wigner-Seitz
unit cell and the first Brillouin zone, respectively, of the monolayer
honeycomb lattice.  Equivalently, both may be taken to be centered on
a point of sixfold rotational symmetry in their corresponding spaces.
In this case the Wigner-Seitz unit cell is bounded by the sides of a
primitive hexagon of side length $d$, as in Fig.\ \ref{model_pattern},
and the corresponding first Brillouin zone is then also a hexagon, now
of side length $4\pi/(3\sqrt{3}d)$, but rotated by $90^{\circ}$ with
respect to the Wigner-Seitz hexagon.  Thus, with respect to an origin
at the center of the hexagon, three of its corners occupy the
positions ${\mathbf K}^{(1)}=4\pi/(3\sqrt{3}d)\hat{x}$,
${\mathbf K}^{(2)}=2\pi(\hat{x}+\sqrt{3}\hat{z})/(3\sqrt{3}d)$, and
${\mathbf K}^{(3)}=2\pi(-\hat{x}+\sqrt{3}\hat{z})/(3\sqrt{3}d)$, with
the remaining corners at positions
${\mathbf K}^{(n+3)}=-{\mathbf K}^{(n)};\;n=1,2,3$.

Classically, the generic stable GS phase with magnetic LRO takes the
form of a coplanar spiral configuration of spins defined in terms of
an ordering wave vector ${\mathbf Q}$, plus an angle $\theta$ that
measures the angle between the two spins in each monolayer unit cell
$l$ at position vector ${\mathbf R}_{l}$.  The two classical spins,
each of length $s (\to \infty)$, in unit cell $l$, are written as
\begin{equation}
{\mathbf s}_{l,\rho}=-s[\cos({\mathbf Q}\cdot{\mathbf R}_{l}+\theta_{\rho})\hat{z}_{s}+\sin({\mathbf Q}\cdot{\mathbf R}_{l}+\theta_{\rho})\hat{x}_{s}]\,; \quad \rho=1,2\,,  \label{classical_spins_eq}
\end{equation}
where the index $\rho$ labels the two sites in the unit cell, and
$\hat{x}_{s}$ and $\hat{z}_{s}$ are two orthogonal unit vectors that define
the spin-space plane, as illustrated in Fig.\ \ref{model_pattern}.  With
no loss of generality, we may choose the two angles $\theta_{\rho}$ such that $\theta_{1}=0$ and $\theta_{2}=\theta$ for the spins on
triangular sublattices 1 and 2, respectively.  

Within this framework the N\'{e}el GS spin configuration shown in
Fig.\ \ref{model_pattern}(b) is now specified by the ordering wave
vector ${\mathbf Q}={\mathbf \Gamma}=(0,0)$, together with the value
$\theta=\pi$ for the relative angle variable between the two sites in
the unit cell.  Similarly, the N\'{e}el-II GS spin configuration shown
in Fig.\ \ref{model_pattern}(c) is specified by the ordering wave
vector ${\mathbf Q}={\mathbf M}^{(2)}=2\pi/(3d)\hat{z}$, together with
the value $\theta=0$.  We note that ${\mathbf M}^{(2)}$ is just the
vector that defines the midpoint of the edge joining the corners
${\mathbf K}^{(2)}$ and ${\mathbf K}^{(3)}$ of the hexagonal first
Brillouin zone described above.  Thus, the other two inequivalent
N\'{e}el-II states have ordering wave vectors that correspond to the
midpoints of the other two non-parallel edges of the hexagonal first
Brillouin zone, and in each case now together with the value
$\theta=\pi$ for the relative angle variable between the two sites on
the monolayer unit cell shown in Fig.\ \ref{model_pattern}(a).  These
may hence be taken as 
${\mathbf Q}={\mathbf M}^{(1)}=\pi(\sqrt{3}\hat{x}+\hat{z})/(3d)$ and
${\mathbf Q}={\mathbf M}^{(3)}=\pi(-\sqrt{3}\hat{x}+\hat{z})/(3d)$, which are,
respectively, the midpoints of the edges joining corners
${\mathbf K}^{(1)}$ and ${\mathbf K}^{(2)}$, and corners ${\mathbf K}^{(3)}$ and ${\mathbf K}^{(4)}$.

Note that, equivalently, we have
$\mathbf{M}^{(1)}=\frac{1}{2}\boldsymbol{\alpha}$,
$\mathbf{M}^{(2)}=\frac{1}{2}\boldsymbol{\beta}$,
$\mathbf{M}^{(3)}=\frac{1}{2}(\boldsymbol{\beta}-\boldsymbol{\alpha})$.
Hence, each of the N\'{e}el-II states corresponds to an ordering wave
vector that equals exactly one half of a reciprocal lattice vector.
While the generic stable classical GS phase is described by the spin
configuration of Eq.\ (\ref{classical_spins_eq}), it is also known
\cite{Villain:1977_ordByDisord} that exceptions occur if the ordering
wave vector ${\mathbf Q}$ takes a value equal to one half or one
quarter of a reciprocal lattice vector
${\mathbf G}_{i}\equiv k_{i}\boldsymbol{\alpha}+l_{i}\boldsymbol{\beta}$,
with $k_{i}, l_{i} \in \mathbb{Z}$, as for the N\'{e}el-II states.  In
this case the classical GS phase is a two-dimensional manifold that
continuously connects the three ${\mathbf Q}={\mathbf M}^{(i)}$ states with
$i=1,2,3$, which now leads to an infinitely degenerate family (IDF) of
non-planar ground states \cite{Fouet:2001_honey}.  As expected, it can
then be shown (and see Ref.\ \cite{Fouet:2001_honey} for details) that
the effect of quantum fluctuations in leading order (i.e., in the
large-$s$ limit using linear spin-wave theory) is to stabilize the
collinear N\'{e}el-II phases from among the IDF family of solutions.

For the classical ($s \to \infty$) $J_{1}$--$J_{2}$ model on the
monolayer honeycomb lattice, one may show that one value of the
ordering wave vector ${\mathbf Q}$ that minimizes the GS energy is
given by
\begin{equation}
{\mathbf Q} = \frac{2}{\sqrt{3}d}\cos^{-1}\left(\frac{1-2\kappa}{4\kappa}\right)\hat{x}\,, \label{Q_eq_E_minimze}
\end{equation}
which should be taken together with the value $\theta=\pi$ for the
relative phase angle.  Clearly, the spiral pitch angle in Eq.\
(\ref{Q_eq_E_minimze}) is physical only when
$\kappa \geq \frac{1}{6}$, and at the boundary $\kappa = \frac{1}{6}$
we have ${\mathbf Q}=0$.  One finds that the N\'{e}el state (with
${\mathbf Q}=0$) is the stable GS phase for all values
$\kappa \leq \frac{1}{6}$, and a spiral state forms the GS phase for
$\frac{1}{6} < \kappa < \infty$.  We note that as $\kappa \to \infty$,
which is the point where the two triangular sublattices of the
honeycomb lattice decouple, the spiral pitch angle takes the value
$\pm \frac{2}{3}\pi$, which is just the expected classical spin
ordering for a triangular lattice.  In this limit the ordering wave
vector ${\mathbf Q}$ of Eq.\ (\ref{Q_eq_E_minimze}) approaches the
value ${\mathbf K}^{(1)}$ of one of the corners of the hexagonal first
Brillouin zone.  Clearly, there are also five other symmetry-related
${\mathbf Q}$ values for the spiral phase that minimize the classical
GS energy in this case, which are obtained by rotations
$\frac{1}{3}n\pi$, with $n=1,2,3,4,5$, of the ${\mathbf Q}$ vector of
Eq.\ (\ref{Q_eq_E_minimze}).

In fact, it can readily be shown
\cite{Rastelli:1979_honey,Fouet:2001_honey,Mulder:2010_honey} that for
all values $\kappa > \frac{1}{6}$ the classical $J_{1}$--$J_{2}$ model
on the monolayer honeycomb lattice actually has an IDF of
incommensurate, planar, spin spiral GS phases in which the spiral wave
vector ${\mathbf Q}$ can point in an arbitrary direction.  For each
value of the frustration parameter in the range
$\frac{1}{6} < \kappa < \frac{1}{2}$ these classically degenerate
solutions form a closed contour around the center,
${\mathbf Q}={\mathbf \Gamma}=(0,0)$, of the hexagonal first Brillouin zone.
Conversely for each value $\kappa > \frac{1}{2}$ the solutions lie on
pairs of closed contours centered on any two of the inequivalent
corners (say ${\mathbf K}^{(1)}$ and ${\mathbf K}^{(2)}$) of the
hexagonal first Brillouin zone.  Precisely at the classical critical
point $\kappa = \frac{1}{2}$, which marks the transition between two
different spiral phases, the contour for the degenerate values of
${\mathbf Q}$ is formed from the hexagon joining the midpoints
$\mathbf{M}^{(n)}$ of the six edges of the hexagonal first Brillouin zone.  It
is easy to see that this boundary can equivalently be taken as a pair
of equilateral triangles centered on ${\mathbf K}^{(1)}$ and
${\mathbf K}^{(2)}$, respectively, for the first of which one of its
sides is the line joining $\mathbf{M}^{(6)}$ and $\mathbf{M}^{(1)}$,
and for the second of which one of its sides is the line joining
$\mathbf{M}^{(1)}$ and $\mathbf{M}^{(2)}$.  At the critical point
$\kappa=\frac{1}{2}$, the value of ${\mathbf Q}$ from Eq.\
(\ref{Q_eq_E_minimze}) is precisely the midpoint of the line joining
the corners $\mathbf{M}^{(1)}$ and $\mathbf{M}^{(6)}$.  We also note
that precisely at this critical point $\kappa = \frac{1}{2}$ the
spiral phases are also degenerate with the collinear N\'{e}el-II
phase.  As $\kappa \to \infty$, the contours collapse to the points
${\mathbf K}^{(1)}$ and ${\mathbf K}^{(2)}$ themselves.

At leading order in spin-wave theory quantum fluctuations have been
shown \cite{Mulder:2010_honey} to lift this otherwise accidental
degeneracy in favor of specific wave vectors that now minimize the GS
energy from among each IDF of states, thereby leading to the
phenomenon of spiral order by disorder
\cite{Villain:1977_ordByDisord,Villain:1980_ordByDisord,Shender:1982_ordByDisord}.
As the frustration parameter $\kappa$ is increased from the value
$\frac{1}{6}$ to the value $\frac{1}{2}$, one solution for this
selected set of values for ${\mathbf Q}$ moves continuously along the
straight line from the point ${\mathbf Q}={\mathbf \Gamma}=(0,0)$ to
the point ${\mathbf Q}=\mathbf{M}^{(2)}$.  As $\kappa$ is then
increased further to values greater than $\frac{1}{2}$, this selected
value for ${\mathbf Q}$ then moves continuously along an edge of the hexagonal 
first Brillouin zone from $\mathbf{M}^{(2)}$ to the corner
${\mathbf K}^{(2)}$.  For all values $\kappa > \frac{1}{6}$, there are
clearly still six symmetry-related degenerate values of ${\mathbf Q}$
that are so selected by quantum fluctuations, with the other five, in
each case, related to those described above by rotations about
${\mathbf \Gamma}=(0,0)$ of $\frac{1}{3}n\pi$, with $n=1,2,3,4,5$.

For the extreme quantum case $s=\frac{1}{2}$, one expects that quantum
fluctuations might well be sufficiently strong as to melt the coplanar
spiral order, in favor of either collinear quasiclassical magnetic
orderings or nonclassical paramagnetic states, over a wide range of
values of $\kappa$ for the $J_{1}$--$J_{2}$ model on the monolayer
honeycomb lattice.  Similarly, since, in general, quantum fluctuations
tend to favor collinear over non-collinear order, one expects that the
critical value $\kappa^{1,>}_{c}$ of the frustration parameter $\kappa$
beyond which N\'{e}el order melts might be larger than the classical
value of $\frac{1}{6}$ for the spin-$\frac{1}{2}$ $J_{1}$--$J_{2}$
model on the monolayer honeycomb lattice.  By now there is a broad
consensus among authors using a wide variety of calculational
techniques (see, e.g., Refs.\
\cite{Reuther:2011_honey,Albuquerque:2011_honey,Mosadeq:2011_honey,Oitmaa:2011_honey,Mezzacapo:2012_honey,Li:2012_honey_full,Bishop:2012_honeyJ1-J2,RFB:2013_hcomb_SDVBC,Zhang:2013_honey,Ganesh:2013_honey_J1J2mod-XXX,Zhu:2013_honey_J1J2mod-XXZ,Gong:2013_J1J2mod-XXX,Yu:2014_honey_J1J2mod})
that both of these conjectures are true.  In particular, the majority of
calculations yield a value for $\kappa^{1,>}_{c}$ in the approximate
range 0.19 to 0.23.  There is then some controversy about whether the
N\'{e}el state is followed immediately by a paramagnetic PVBC state or
whether there is a small intermediate QSL phase.  Beyond the region of
stability of the PVBC phase many calculations concur that the classical spin
spiral states are still destabilized for a further range of values of
$\kappa$ in favor of a stable GS phase that is either a paramagnetic
SDVBC state or a quasiclassical N\'{e}el-II AFM state (or, indeed,
different regimes of both).  Since both the SDVBC and N\'{e}el-II
states break the same lattice symmetries, many calculations find them
very difficult to differentiate cleanly.  This is particularly true
for methods (such as exact diagonalization and density-matrix
renormalization group techniques) that are based intrinsically on
finite-size lattices, and which need to be extrapolated to the
infinite-lattice ($N \to \infty$) limit.

One of our aims in the present paper is to shed more light on the stability of both possible quasiclassical AFM phases of the spin-$\frac{1}{2}$ $J_{1}$--$J_{2}$
model on the monolayer honeycomb lattice, viz., the N\'{e}el and N\'{e}el-II phases.  In order to do so we now consider the larger spin-$\frac{1}{2}$ $J_{1}$--$J_{2}$--$J_{1}^{\perp}$ model on the bilayer honeycomb lattice, and consider the realms of stability of both phases in the $\kappa\delta$ half-plane with $\kappa \equiv J_{2}/J_{1}>0$ (and $J_{1}>0$), and for arbitrary values of the interlayer coupling parameter, $\delta \equiv J_{1}^{\perp}/J_{1}$.  While we consider both signs of $\delta$, we will principally be interested in the case of an AFM interlayer coupling ($J_{1}^{\perp}>0$), such that NN spins between the two layers of the $AA$-stacked bilayer are antiparallel to one another.  Thus, it is important to note from the outset that although we will also consider cases with $J_{1}^{\perp}<0$, where physically the NN spins between the two layers would energetically prefer to be parallel to one another, we consider here only the stability of the two phases with each monolayer having either N\'{e}el or N\'{e}el-II ordering, but with the two layers connected so that NN spins between them are antiparallel.  Obviously these bilayer phases are unphysical when $\delta < 0$.  Nevertheless, we study them also in this unphysical regime, as well as in the physical regime where $\delta > 0$, since by doing so we can shed particular light on the stability of the two collinear monolayer phases (i.e., when $\delta=0$).   As a foretaste of our results, we remark that we will demonstrate rather clearly why both phase boundaries for the monolayer are difficult to calculate with high accuracy, by showing how sensitive the boundaries in the bilayer model are to small changes in $\delta$ near the monolayer limit, $\delta=0$.

\section{The coupled cluster method}
\label{ccm_sec}
The CCM provides an accurate and versatile technique of {\it ab
  initio} quantum many-body theory, which has been applied with
considerable success in a wide range of physical and chemical contexts
(and see, e.g., Refs.\
\cite{Kummel:1978_ccm,Bishop:1978_ccm,Bishop:1982_ccm,Bishop:1987_ccm,Bartlett:1989_ccm,Bishop:1991_TheorChimActa_QMBT,Bishop:1998_QMBT_coll,Fa:2004_QM-coll,Bartlett:2007_ccm}).
Very importantly, the method is systematically improvable within
several well-defined hierarchical approximation schemes that are
guaranteed to approach the exact results in the limit $n \to \infty$,
where $n$ is an index that signifies the order of the approximation
within some specified scheme.  Of course, computational considerations
generally restrict one in practice to the highest values $n$ in the
sequence that are attainable, and one then needs to extrapolate the
partial sequences of values obtained for any GS or excited-state (ES)
parameter to the limit $n \to \infty$, as we describe below.

It is important to realize from the outset that this extrapolation to
the exact physical limit ($n \to \infty$) is the {\it only}
approximation that ever needs to be made when implementing the CCM in
practice.  In particular, since the method is both size-extensive and
size-consistent at every approximation level $n$, it can be
implemented from the very beginning in the infinite system
($N \to \infty$) limit.  This immediately removes the need for any
finite-size scaling of the sort that is required by most alternative methods
such as, for example, the exact diagonalization (ED) of small lattices
and density-matrix renormalization group (DMRG) techniques.  This
additional associated source of errors is thus circumvented by use of
the CCM.

In addition to these obvious advantages the CCM has two other
important attributes.  Thus, at every $n$th-order level of
approximation, the CCM also exactly preserves both the
Hellmann-Feynman theorem and the Goldstone linked cluster theorem.
These attributes are key to understanding the success of the method in
providing results for a variety of both GS and ES parameters for the system being
studied that are both highly accurate and also self-consistent.  By
now these include a large number of spin-lattice systems in quantum
magnetism, and accordingly we refer the reader to the extensive
literature (and see, e.g., Refs.\
\cite{DJJFarnell:2014_archimedeanLatt,Li:2012_honey_full,Bishop:2012_honeyJ1-J2,RFB:2013_hcomb_SDVBC,Bishop:2017_honeycomb_bilayer_J1J2J1perp,Li:2017_honeycomb_bilayer_J1J2J1perp_Ext-M-Field,Zeng:1998_SqLatt_TrianLatt,Fa:2004_QM-coll,DJJF:2011_honeycomb,PHYLi:2012_honeycomb_J1neg,Bishop:2012_honey_circle-phase,Bishop:2014_honey_XY,Li:2014_honey_XXZ,Bishop:2014_honey_XXZ_nmp14,Bishop:2015_honey_low-E-param,Bishop:2016_honey_grtSpins,Li:2016_honeyJ1-J2_s1,Li:2016_honey_grtSpins}
and references cited therein) for complete details.  Nevertheless, we
present below a brief recapitulation of those features of the method
that are most germane to the present analysis.

In order to initiate any application of the CCM one needs first to
choose some (one or more) suitable normalized model (or reference)
state $|\Phi\rangle$, on top of which the quantum correlations present
in the exact GS or ES wave functions appropriate to the phase of the
system under study are then (in principle exactly) incorporated in
terms of correlations operators that involve a very specific
exponentiated form, which is one of the distinguishing key features of
the method.  It is then these correlations operators that are
systematically approximated to higher and higher orders, as discussed
above, and to which we return in more detail below.  In order to
calculate the stability regimes of the GS phases discussed here, we
utilize (separately) both the quasiclassical AFM states (viz., the
N\'{e}el and N\'{e}el-II states, as illustrated in Figs.\
\ref{model_pattern}(b) and \ref{model_pattern}(c), respectively) for
each honeycomb-lattice monolayer, as our two choices for CCM model
state.  Both of the quasiclassical AFM states are eigenstates of the
operator $s_{T}^{z}$, where
$s_{T}^{z} \equiv \sum_{k,\alpha}s_{k,\alpha}^{z}$ is the total $z$
component of spin for the system as a whole, using global spin axes,
with $s_{T}^{z}=0$.  For each collinear phase we also present results
for the excitation energy $\Delta$ from the corresponding
$s_{T}^{z}=0$ ground state to the lowest-lying excited state in the
$|s_{T}^{z}|=1$ sector.

Once the choice of model state has been made the only remaining
decision before the CCM can be implemented computationally is the
choice of approximation scheme to use.  This simply involves which
multispin-flip configurations that are to be retained in the GS and ES
ket- and bra-state correlation operators that are used to parametrize
the corresponding exact GS and ES wave functions.  We use here the
very well-tested and much used scheme known as the localized
lattice-animal-based subsystem (LSUB$n$) hierarchy.  At the $n$th
order in the LSUB$n$ scheme those multispin-flip configurations to be
retained in the CCM correlation operators are defined to be those that
describe clusters of spins that span a range of $n$ or fewer
contiguous sites on the lattice.  A set of lattice sites is defined to
be contiguous in this sense if every site of the set is a NN to at
least one other member of the set.  Equivalently, in the terminology
of graph theory, the LSUB$n$ approximation retains all multispin-flip
configurations defined on lattice animals (or polyominos) of size $n$
or smaller.  We define a single spin-flip on site $l$ and layer
$\alpha$ as requiring the action of the spin-raising operator
$s_{l,\alpha}^{+}\equiv s_{l,\alpha}^{x}+{\mathrm i}s_{l,\alpha}^{y}$ acting
once on the model ket state, which is now described in the very
convenient (and universal) local set of spin axes in which a different passive
rotation has been made at every site so that every spin points downwards (i.e.,
along the negative $z_{s}$ axis).

It is clear that as the truncation index $n \to \infty$ the LSUB$n$
approximation becomes exact.  We use the space- and point-group
symmetries of both the system Hamiltonian and the particular CCM model
state being employed to reduce the set of {\it independent} multi-spin
configurations retained within a given LSUB$n$ approximation to the
minimal number $N_{f}=N_{f}(n)$.  For example, for our system the
operator $s_{T}^{z}$ defined above is conserved, and we have that both
quasiclassical AFM states lie in the sector with $s_{T}^{z}=0$.
Hence, only GS multispin-flip configurations are retained that are in
accord with $s_{T}^{z}=0$.  Similarly, for the calculation of the
excitation energy $\Delta$ to the lowest-lying state in the $s_{T}^{z}=1$ conserved sector,  we only retain ES multispin-flip configurations that
satisfy $s_{T}^{z}=1$.  In both cases the LSUB$n$ hierarchy is used
to order the approximations.  The precise way that the GS and ES correlation operators are defined is give in Ref.\ \cite{Bishop:2017_honeycomb_bilayer_J1J2J1perp}.

Nevertheless, this resulting number $N_{f}(n)$ of
fundamental configurations typically increases rapidly as a function
of the truncation index $n$, and the available computing power limits
us in practice to a maximum value, $n_{\rm max}$.  For example, for
the present spin-$\frac{1}{2}$ model on the bilayer honeycomb lattice,
even by making use of massively parallel supercomputing resources,
both to derive (using a specially tailored computer algebra package
\cite{ccm_code}) and solve \cite{Zeng:1998_SqLatt_TrianLatt} the sets of
$N_{f}$ coupled CCM bra- and ket-state equations for both GS and ES
parameters, we are constrained to $n_{\rm max}=10$.  Thus, we have
$N_{f}(10)=70\,118$ $(197\,756)$ for the calculations of GS quantities
for the present model using the N\'{e}el (N\'{e}el-II) states on each
monolayer as the CCM model state, respectively.  The corresponding
numbers of fundamental configurations retained at the LSUB10 level of
approximation for the calculations of the excitation energy $\Delta$ from the $s_{T}^{z}=0$ N\'{e}el (N\'{e}el-II) ground states to the lowest-lying excited state in the respective $s_{T}^{z}=1$ sectors are $N_{f}(10)=121\,103$
$(352\,779)$, respectively.

In Sec.\ \ref{results_sec} we present LSUB$n$ results for both the GS
magnetic order parameter (i.e., the average local on-site
magnetization) $M$ and the excitation energy $\Delta$ to the lowest-lying state in the $s_{T}^{z}=1$ sector, based on the
use of both the N\'{e}el and N\'{e}el-II states on each monolayer as
the CCM model state.  Specifically, we define
\begin{equation}
M \equiv -\frac{1}{N}\langle\tilde{\Psi}|\sum_{k,\alpha}s_{k,\alpha}^{z}|\Psi\rangle\,,  \label{M_Eq}
\end{equation}
in the local rotated spin coordinate frames described above in which
each spin points along the negative $z_{s}$ axis, where
$\langle\tilde{\Psi}|$ and $|\Psi\rangle$ are the GS bra and ket
many-body wave functions (normalized so that
$\langle\tilde{\Psi}|\Psi\rangle=1$), here parametrized in the usual
CCM fashion (and see, e.g., Ref.\
\cite{Bishop:2017_honeycomb_bilayer_J1J2J1perp} for details).  Unlike
in many other methods, such as QMC simulations, where quasiclassical
order is naturally investigated and characterized through the
parameter $M^{2}$, it is both possible and natural within the CCM
framework to calculate the parameter $M$ directly and hence to use it
to quantify the degree of order.  This also has the additional
advantage of showing rather clearly points where the phase under
consideration becomes unstable, namely where $M$ goes to zero and then
becomes negative.

As we have noted previously, the sole approximation that now needs to
be made is to extrapolate the LSUB$n$ approximants $M(n)$ and
$\Delta(n)$ for $M$ and $\Delta$, respectively, to the exact limit,
$n \to \infty$.  By now, a great deal of practical experience, from
applications to many different spin-lattice models, has shown that the
consistent use of simple extrapolation schemes for various physical
parameters always leads to accurate results.  Thus, for spin-lattice
models with a high degree of frustration present, particularly in
situations where the system is close to a QPT or where the order
parameter $M$ is either zero or very close to zero, the appropriate
extrapolation scheme for $M$ (and see, e.g., Refs.\
\cite{Li:2012_honey_full,Bishop:2012_honeyJ1-J2,RFB:2013_hcomb_SDVBC,Bishop:2017_honeycomb_bilayer_J1J2J1perp,Li:2017_honeycomb_bilayer_J1J2J1perp_Ext-M-Field,DJJF:2011_honeycomb,PHYLi:2012_honeycomb_J1neg,Bishop:2012_honey_circle-phase,Bishop:2014_honey_XY,Li:2014_honey_XXZ,Bishop:2014_honey_XXZ_nmp14,Li:2016_honeyJ1-J2_s1,Li:2016_honey_grtSpins,Li:2018_crossStripe_low-E-param})
is given as
\begin{equation}
M(n) = \mu_{0}+\mu_{1}n^{-1/2}+\mu_{2}n^{-3/2}\,,   \label{M_extrapo_frustrated}
\end{equation}
from fits with LSUB$n$ data sets to which we extract the LSUB$\infty$
extrapolant $\mu_{0}$ for $M$.  The appropriate scheme for the excitation energy $\Delta$ to the lowest-lying state in the $s_{T}^{z}=1$ sector is found (see e.g., Refs.\
\cite{Bishop:2017_honeycomb_bilayer_J1J2J1perp,Bishop:2015_honey_low-E-param,Li:2016_honeyJ1-J2_s1,Li:2018_crossStripe_low-E-param,Kruger:2000_JJprime,Richter:2015_ccm_J1J2sq_spinGap,Bishop:2015_J1J2-triang_spinGap})
to be given as
\begin{equation}
\Delta(n) = d_{0}+d_{1}n^{-1}+d_{2}n^{-2}\,.   \label{Eq_spin_gap}
\end{equation}
Once again, by fitting with LSUB$n$ data sets, we can extract the
corresponding LSUB$\infty$ extrapolant $d_{0}$ for $\Delta$.

There is one additional point that deserves to be mentioned in
connection with the use of extrapolation schemes, such as those in
Eqs.\ (\ref{M_extrapo_frustrated}) or (\ref{Eq_spin_gap}), in
practice.  This involves the possible presence of so-called
``staggering effects'' in the sequences of approximants.  A well-known
example occurs in perturbation theory where {\it exact} extrapolation
schemes for various physical quantities are often known, but where the
even and odd sequences of approximants from $n$th-order perturbation
theory (i.e., those with $n=2m$ and those with $n=2m-1$, respectively,
where $m \in \mathbb{Z}^{+}$ is a positive integer) involve an
additional staggering effect.  In this case both sequences obey an
extrapolation scheme of the same sort (i.e., with the same leading
exponent), but where the coefficients (other than the leading constant
term, corresponding to the exact, $n \to \infty$, limit) are not
identical.  Clearly, one should not then mix even and odd terms
together in a single extrapolation scheme, unless the staggering is
also incorporated explicitly.  Such an explicit inclusion of the
staggering is always difficult to achieve in a robust manner.  In
practice it is almost always circumvented by extrapolating only the
even-order (or only the odd-order) terms.  A similar odd/even [i.e.,
$(2m-1)/2m$] staggering is also always present to a greater or lesser
degree for all LSUB$n$ sequences of approximants.  It is for that
reason that we restrict attention here in Sec.\ \ref{results_sec}
only to even-order LSUB$n$ approximations (i.e., those with $n=2m$).
In principle, we could also separately explore the odd-order LSUB$n$
approximants.  However, since the Hamiltonian of Eq.\ (\ref{H_eq})
contains only terms that are bilinear in the spin operators, it is
much more natural in this case to restrict attention to the even-order
approximations.

It has been noted previously
\cite{Bishop:2012_honeyJ1-J2,RFB:2013_hcomb_SDVBC,Li:2016_honeyJ1-J2_s1,Bishop:2017_honeycomb_bilayer_J1J2J3J1perp}
that, while the $(2m-1)/2m$ staggering in LSUB$n$ sequences of
approximants is common to all spin lattices, a further subtlety arises
in the case of honeycomb-lattice models.  For such models one observes
an {\it additional} staggering effect, such that in the even-order
series of LSUB$n$ approximants for some observable quantities the terms
with $n=(4m-2)$ are offset (or staggered) with respect to those with
$n=4m$.  As has been pointed out elsewhere
\cite{Li:2016_honeyJ1-J2_s1}, it is likely that this additional
$(4m-2)/4m$ staggering effect arises from the non-Bravais nature of
the honeycomb lattice.  Thus, each of the two interlocking triangular
Bravais sublattices, which comprise the honeycomb lattice, exhibits a
$(2m-1)/2m$ staggering of the usual kind.  In turn, this then leads to
the ``doubling'' of the effect in the composite honeycomb lattice,
where it manifests itself as the observed $(4m-2)/4m$ staggering.  In
order to take this additional effect into account, and since we are
restricted computationally to performing LSUB$n$ calculations for the
present model to those with $n \leq 10$, most of the extrapolations
discussed in Sec.\ \ref{results_sec} are based on the LSUB$n$ data
sets with $n=\{2,6,10\}$.

\begin{figure*}[t]
\begin{center}
\mbox{
\hspace{-1.0cm}
\subfigure[]{\includegraphics[width=6.1cm]{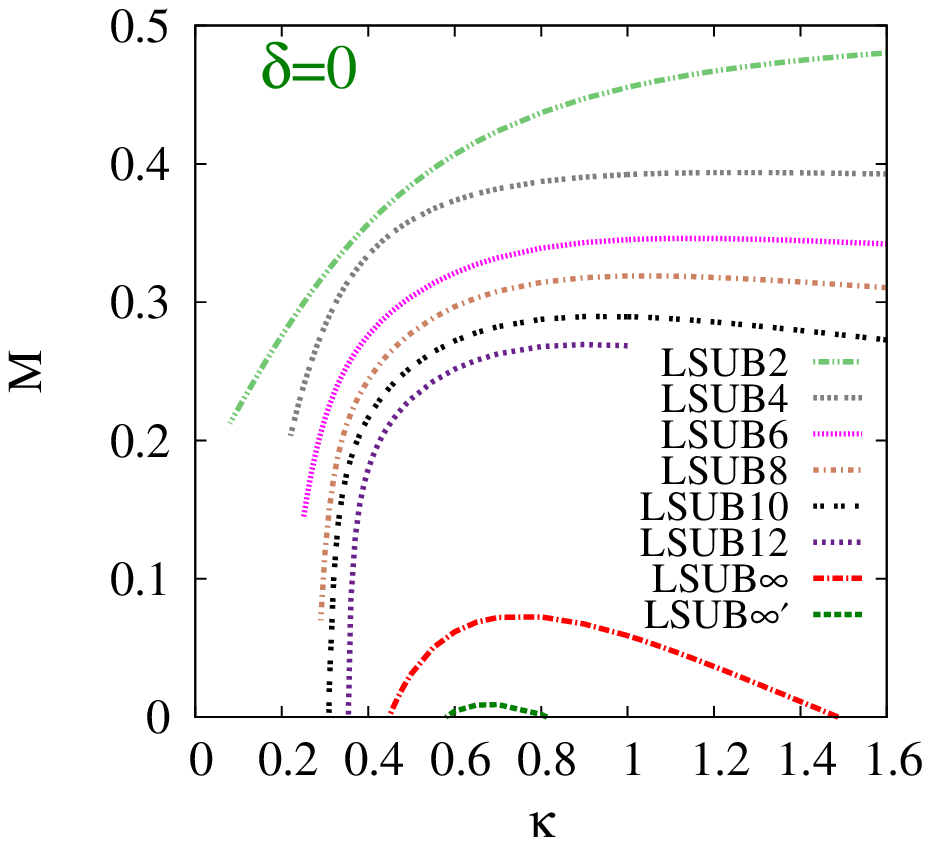}}
\hspace{-1.7cm}
\subfigure[]{\includegraphics[width=6.1cm]{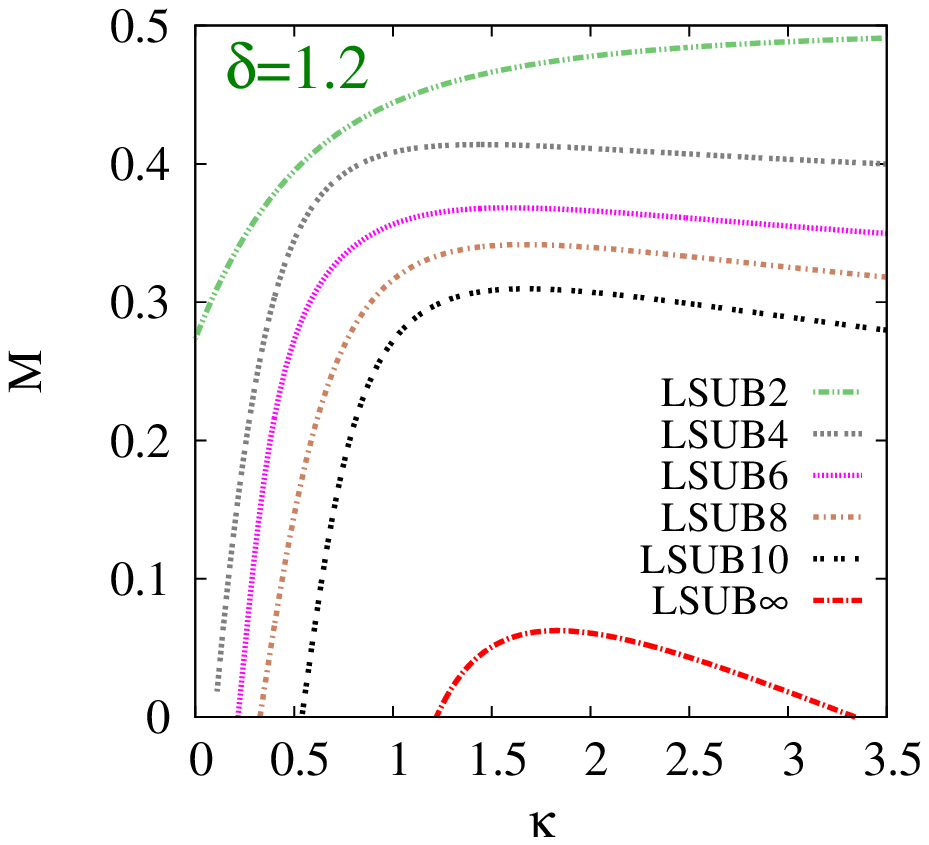}}
\hspace{-1.7cm}
\subfigure[]{\includegraphics[width=6.1cm]{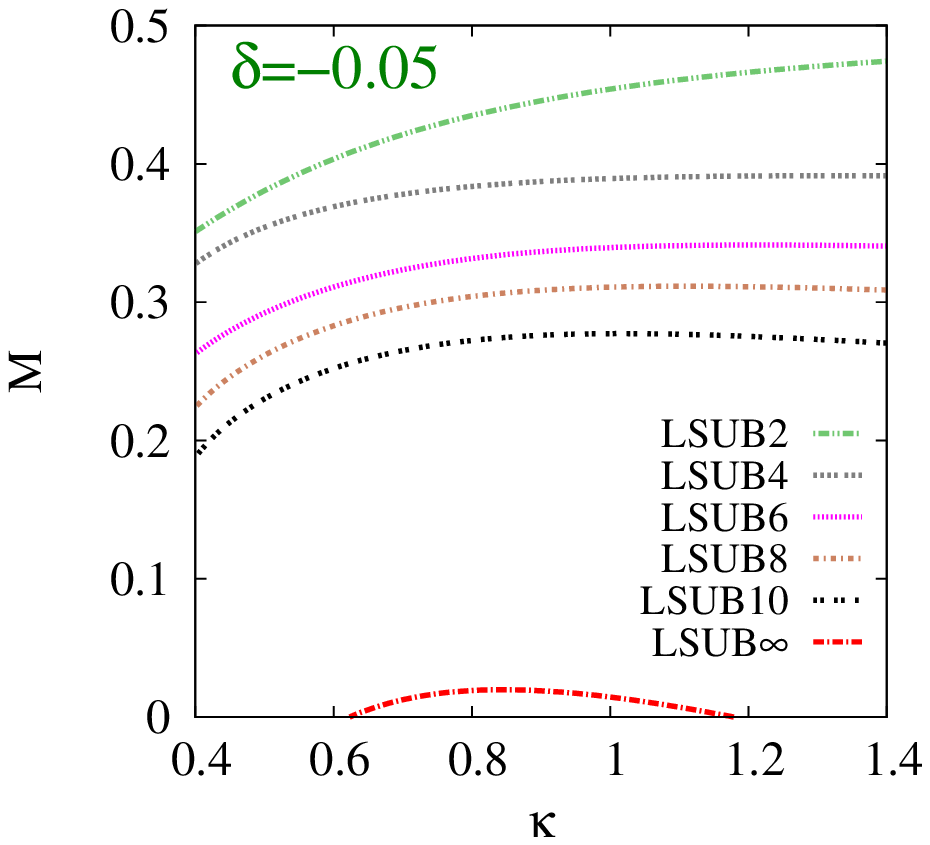}}
}
\caption{CCM results for the GS magnetic order parameter $M$ versus
  the intralayer frustration parameter, $\kappa \equiv J_{2}/J_{1}$,
  for the spin-$\frac{1}{2}$ $J_{1}$--$J_{2}$--$J_{1}^{\perp}$ model
  on the bilayer honeycomb lattice (with $J_{1}>0$), for three
  selected values of the scaled interlayer exchange coupling constant,
  $\delta \equiv J_{1}^{\perp}/J_{1}$: (a) $\delta=0$, (b)
  $\delta=1.2$, and (c) $\delta=-0.05$.  Results based on the
  N\'{e}el-II state on each monolayer (and the two layers coupled so that NN spins between them are antiparallel to one another, even in case (c) where $\delta < 0$) as CCM model state are shown in
  LSUB$n$ approximations with $n=2,4,6,8$, and $10$, (and also with
  $n=12$ for the special case $\delta=0$ of the $J_{1}$--$J_{2}$
  monolayer), together with the LSUB$\infty$ extrapolated results
  based on Eq.\ (\ref{M_extrapo_frustrated}) and the LSUB$n$ data sets
  $n=\{2,6,10\}$.  For the case $\delta=0$ only we also show the
  corresponding LSUB$\infty'$ extrapolation based on the LSUB$n$ data set
  $n=\{4,8,12\}$.}
\label{M_raw_extrapo_fix-J1perp_neel-II}
\end{center}
\end{figure*}  

\section{Results}
\label{results_sec}
Since the stability of the N\'{e}el phase has been discussed by us
previously \cite{Bishop:2017_honeycomb_bilayer_J1J2J1perp} in the
sector where $\kappa > 0$ and $\delta > 0$, we concentrate attention
initially on the N\'{e}el-II phase.  In Fig.\
\ref{M_raw_extrapo_fix-J1perp_neel-II} we first show results for the
magnetic order parameter $M$ as a function of the intralayer
frustration parameter $\kappa \equiv J_{2}/J_{1}$ for three separate
fixed values of the interlayer coupling parameter,
$\delta=0,1.2,-0.05$.  In each case the CCM model state comprises the N\'{e}el-II state on each
monolayer and the two layers connected so that NN spins between the layers are antiparallel to one another (even in the case with $\delta < 0$, where such interlayer AFM coupling is unphysical in the sense that the corresponding state with NN interlayer spins parallel to one another is clearly energetically favored).  Results are shown in each case at
LSUB$n$ levels of approximation with $n=2,4,6,8$, and $10$.  For the
special case $\delta=0$ alone, shown in Fig.\
\ref{M_raw_extrapo_fix-J1perp_neel-II}(a), which corresponds to the
$J_{1}$--$J_{2}$ model on the honeycomb-lattice monolayer, we also
show LSUB12 results, since these are computationally feasible in this
case (but not, as we have indicated previously, for the coupled
bilayer cases with $\delta \neq 0$).  Each of the cases shown in Fig.\
\ref{M_raw_extrapo_fix-J1perp_neel-II} clearly illustrates the $(4m-2)/4m$
staggering effect of the LSUB$n$ sequences of approximations that we
discussed in Sec.\ \ref{ccm_sec}.  For that reason we restrict
ourselves to showing LSUB$\infty$ extrapolations of our LSUB$n$ results
in the general case, when $\delta \neq 0$, which are based on Eq.\
(\ref{M_extrapo_frustrated}) and which use the LSUB$n$ data sets with
$n=\{2,6,10\}$ as input.

The results shown in Fig.\ \ref{M_raw_extrapo_fix-J1perp_neel-II}
demonstrate very clearly that for each of the values of $\delta$
displayed there exist lower and upper critical points
$\kappa^{2,<}_{c}(\delta)$ and $\kappa^{2,>}_{c}(\delta)$, such that
N\'{e}el-II order exists on each of the coupled monolayers, for a given value of $\delta$, only for
values of $\kappa$ in the range
$\kappa^{2,<}(\delta)<\kappa<\kappa^{2,>}_{c}(\delta)$.  For example,
for the $J_{1}$--$J_{2}$ monolayer $(\delta=0)$ shown in Fig.\
\ref{M_raw_extrapo_fix-J1perp_neel-II}(a), we have
$\kappa^{2,<}_{c}(0)\approx 0.45$ and
$\kappa^{2,>}_{c}(0) \approx 1.49$ from the LSUB$\infty$ extrapolation
using the LSUB$n$ data set $n=\{2,6,10\}$.  The corresponding values
from the LSUB$\infty'$ extrapolation using LSUB$n$ data with
$n=\{4,8,12\}$ are $\kappa^{2,<}_{c}(0) \approx 0.58$ and
$\kappa^{2,>}_{c}(0) \approx 0.81$.  While the differences between the
two extrapolations may appear somewhat large, we emphasize now that
these are {\it not} indicative of our overall errors.  Rather, they
arise from a specific (and wholly natural and completely unavoidable)
region of great sensitivity near to the line $\delta=0$ in the
$\kappa \delta$ plane, as we explain more fully below.

It is already apparent from Fig.\
\ref{M_raw_extrapo_fix-J1perp_neel-II}(a) that the N\'{e}el-II order
in the honeycomb-lattice $J_{1}$--$J_{2}$ monolayer is quite fragile,
with values of the order parameter $M < 0.1$ over the whole range of
values for $\kappa$ for the values of $\delta$ shown.  However, we find that as $\delta$ is first
increased from zero, the interlayer coupling acts to stabilize the
N\'{e}el-II phase rather rapidly, so that by the time $\delta = 0.3$
the maximum value of the order parameter is about 0.2 (at a value of
$\kappa$ around 0.7).  In this region the values of
$\kappa^{2,<}_{c}(\delta)$ are slightly less than
$\kappa^{2,<}_{c}(0)$, thus exhibiting a reentrant behavior, while the
values of $\kappa^{2,>}_{c}(\delta)$ grow monotonically with $\delta$
and are appreciably greater than $\kappa^{2,>}_{c}(0)$.  As $\delta$
is increased further, however, the interlayer coupling now acts to
reduce the N\'{e}el-II order, although at the same time the range of
the values of the frustration parameter $\kappa$ over which it exists
tends to {\it increase} at first.  Thus, while
$\kappa^{2,<}_{c}(\delta)$ and $\kappa^{2,>}_{c}(\delta)$ now both
increase, the latter increases at a faster rate than the former, at
least initially.  We find that there exist values
$\delta = \delta^{l}_{2}$ $(\approx 0.2)$, and
$\delta = \delta^{u}_{2}$ $(\approx 1.2)$ at which, respectively,
$\kappa^{2,<}_{c}(\delta^{l}_{2}) = \kappa^{{\rm min}}_{2} \approx
0.35$, and
$\kappa^{2,>}_{c}(\delta^{u}_{2}) = \kappa^{{\rm max}}_{2} \approx
3.34$, such that for all values of $\delta$ we have
$\kappa^{2,<}_{c}(\delta) \geq \kappa^{{\rm min}}_{2}$ and
$\kappa^{2,>}_{c}(\delta) \leq \kappa^{{\rm max}}_{2}$.  Figure
\ref{M_raw_extrapo_fix-J1perp_neel-II}(b) displays the results for $M$
at a value $\delta =1.2 \approx \delta^{u}_{2}$, from which we see
that $\kappa^{{\rm max}}_{2} \approx 3.34$.  As $\delta$ is now
increased beyond $\delta^{u}_{2}$, we find that
$\kappa^{2,<}_{c}(\delta)$ continues to increase while
$\kappa^{2,>}_{c}(\delta)$ now decreases.  Finally, when $\delta$
attains the value $\delta^{{\rm max}}_{2}$, the two critical values
merge,
$\kappa^{2,<}_{c}(\delta^{{\rm
    max}}_{2})=\kappa^{2,>}_{c}(\delta^{{\rm max}}_{2})\equiv
\kappa^{u}_{2}$, such that for all values
$\delta > \delta^{{\rm max}}_{2}$ N\'{e}el-II order is absent,
whatever the value of the frustration parameter $\kappa$.

Figure \ref{M_raw_extrapo_fix-J1perp_neel-II}(c) displays the effect of
introducing a weak ferromagnetic interlayer coupling $(\delta < 0)$
between the two monolayers with N\'{e}el-II ordering.  Clearly, the
already rather fragile LRO in each monolayer is now weakened further.
The two critical values $\kappa^{2,<}_{c}(\delta)$ and
$\kappa^{2,>}_{c}(\delta)$ move closer together as $\delta$ is made
more negative, until $\delta$ reaches a value
$\delta = \delta^{{\rm min}}$, at which value we have
$\kappa^{2,<}_{c}(\delta^{{\rm
    min}}_{2})=\kappa^{2,>}_{c}(\delta^{{\rm min}}_{2})\equiv
\kappa^{l}_{2}$.  N\'{e}el-II order is then absent for all values
$\delta < \delta^{{\rm min}}_{2}$, for any value of the frustration
parameter $\kappa$.  Based on the LSUB$\infty$ extrapolations,
which use Eq.\ (\ref{M_extrapo_frustrated}) together with the
LSUB$n$ data sets with
$n=\{2,6,10\}$ as input, we find the values
$\delta^{{\rm min}}_{2} \approx -0.06$ and
$\kappa^{l}_{2} \approx 0.9$, together with the corresponding values at the upper boundary of
N\'{e}el-II order, $\delta^{{\rm max}}_{2} \approx
1.51$ and $\kappa^{u}_{2} \approx 2.7$.

\begin{figure*}[t]
\begin{center}
\mbox{
\hspace{-1.0cm}
\subfigure[]{\includegraphics[width=6.1cm]{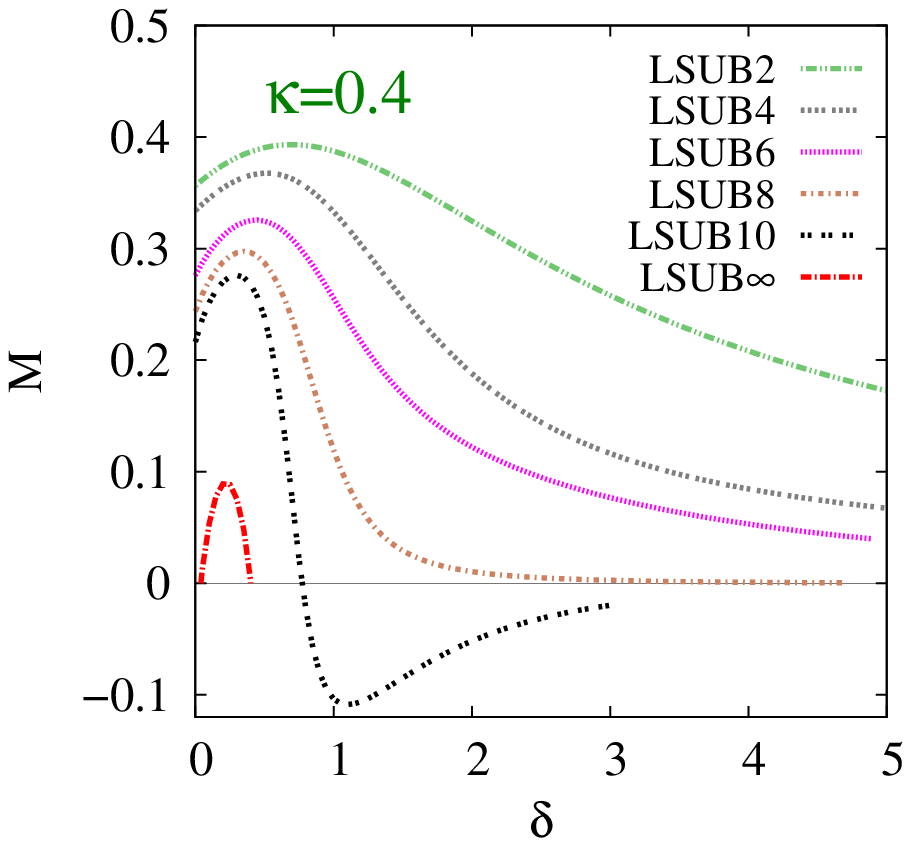}}
\hspace{-1.7cm}
\subfigure[]{\includegraphics[width=6.1cm]{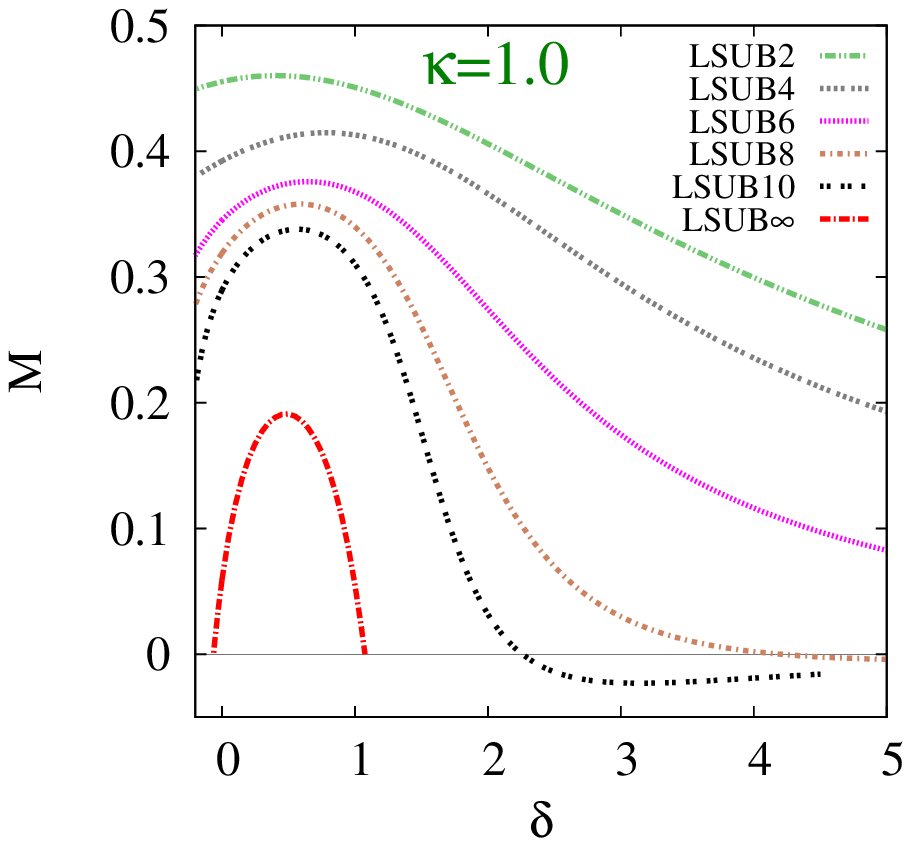}}
\hspace{-1.7cm}
\subfigure[]{\includegraphics[width=6.1cm]{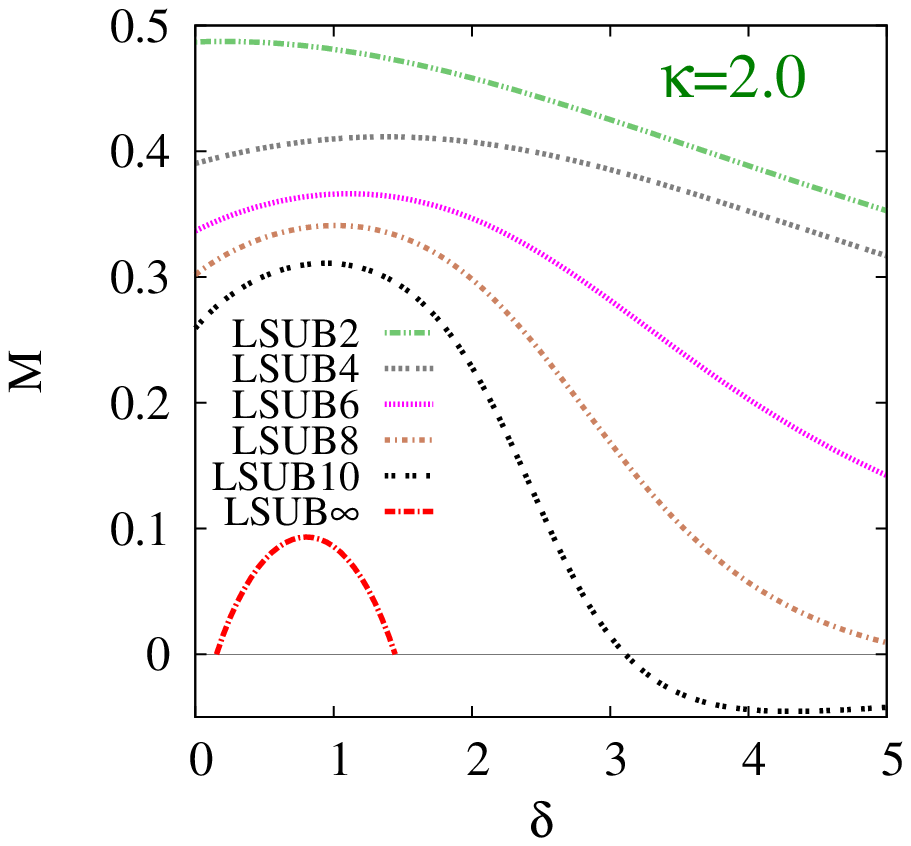}}
}
\caption{CCM results for the GS magnetic order parameter $M$ versus
  the scaled interlayer exchange coupling constant,
  $\delta \equiv J_{1}^{\perp}/J_{1}$, for the spin-$\frac{1}{2}$
  $J_{1}$--$J_{2}$--$J_{1}^{\perp}$ model on the bilayer honeycomb
  lattice (with $J_{1}>0$), for three selected values of the
  intralayer frustration parameter, $\kappa \equiv J_{2}/J_{1}$: (a)
  $\kappa=0.4$, (b) $\kappa=1.0$, and (c) $\kappa=2.0$.  Results based
  on the N\'{e}el-II state on each monolayer (and the two layers coupled so that NN spins between them are antiparallel to one another) as CCM model state are shown in LSUB$n$
  approximations with $n=2,4,6,8$, and $10$, together with the corresponding
  LSUB$\infty$ extrapolated result based on Eq.\
  (\ref{M_extrapo_frustrated}) and the LSUB$n$ data sets 
  $n=\{2,6,10\}$.}
\label{M_raw_extrapo_fix-J2_neel-II}
\end{center}
\end{figure*}

In Fig.\ \ref{M_raw_extrapo_fix-J2_neel-II} we exhibit the effect of
the interlayer coupling on the N\'{e}el-II order in a different way by
showing $M$ versus $\delta$ curves for three separate values of the
intralayer frustration parameter, $\kappa \equiv J_{2}/J_{1}$.  As can
be clearly seen from each of Figs.\
\ref{M_raw_extrapo_fix-J2_neel-II}(a),
\ref{M_raw_extrapo_fix-J2_neel-II}(b), and
\ref{M_raw_extrapo_fix-J2_neel-II}(c), the initial effect of the
interlayer AFM NN coupling, as $\delta$ is increased from zero, is to
increase the order parameter $M$, thereby enhancing the stability of
the N\'{e}el-II ordering on each monolayer.  In each case the effect
reaches a maximum at a certain value of $\delta$, which depends on the
specific value chosen for $\kappa$.  Increasing $\delta$ further then
reduce the N\'{e}el-II order, until (in the extrapolated LSUB$\infty$
limit) an upper critical value $\delta^{c,>}_{2}(\kappa)$ is reached,
beyond which (for a given value of $\kappa$) N\'{e}el-II order
disappears.  Similarly, in each case there is a lower critical value
$\delta^{c,<}_{2}(\kappa)$, below which N\'{e}el-II is wholly absent.
\begin{figure}[t]
\begin{center}
  \includegraphics[width=10cm]{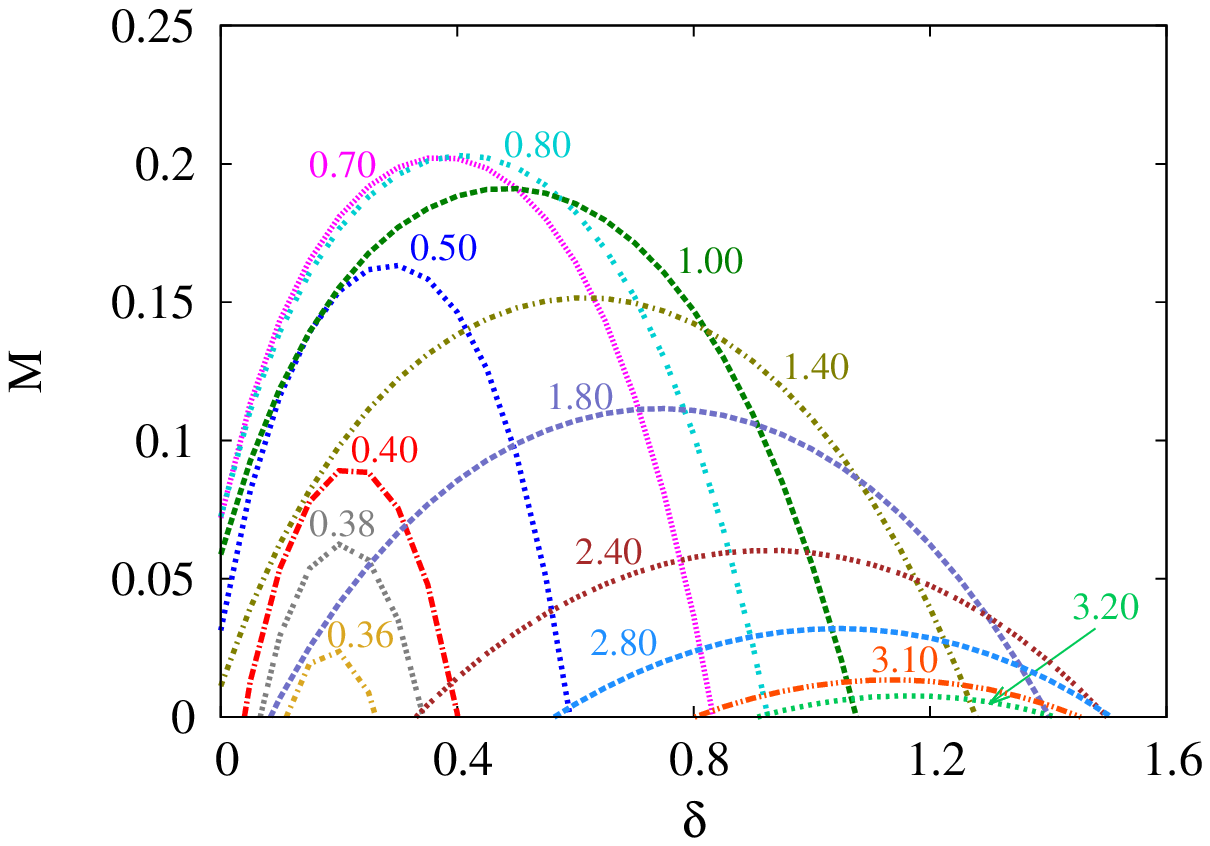}
  \caption{CCM results for the GS magnetic order parameter $M$ versus
    the scaled interlayer exchange coupling constant,
    $\delta \equiv J_{1}^{\perp}/J_{1}$, for the spin-$\frac{1}{2}$
    $J_{1}$--$J_{2}$--$J_{1}^{\perp}$ model on the bilayer honeycomb
    lattice (with $J_{1}>0$), for a variety of values of the
    intralayer frustration parameter, $\kappa \equiv J_{2}/J_{1}$.
  In each case we show extrapolated results, based on the N\'{e}el-II
  state on each monolayer (and the two layers coupled so that NN spins between them are antiparallel to one another) as CCM model state, obtained from using Eq.\
  (\ref{M_extrapo_frustrated}) with the corresponding LSUB$n$ data
  sets $n=\{2,6,10\}$.}
\end{center}
\label{M_J2fix-selective_extrapo_neel-II}
\end{figure}

\begin{figure*}
\begin{center}
\mbox{
\hspace{-1.0cm}
\subfigure[]{\includegraphics[width=6.1cm]{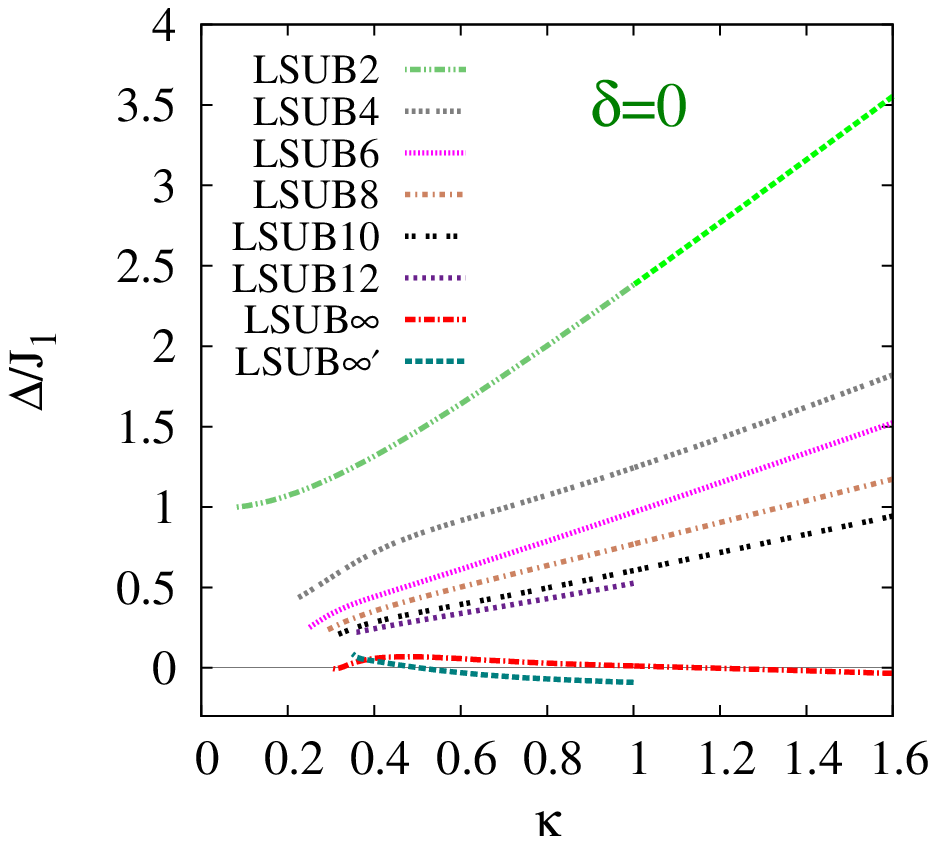}}
\hspace{-1.7cm}
\subfigure[]{\includegraphics[width=6.1cm]{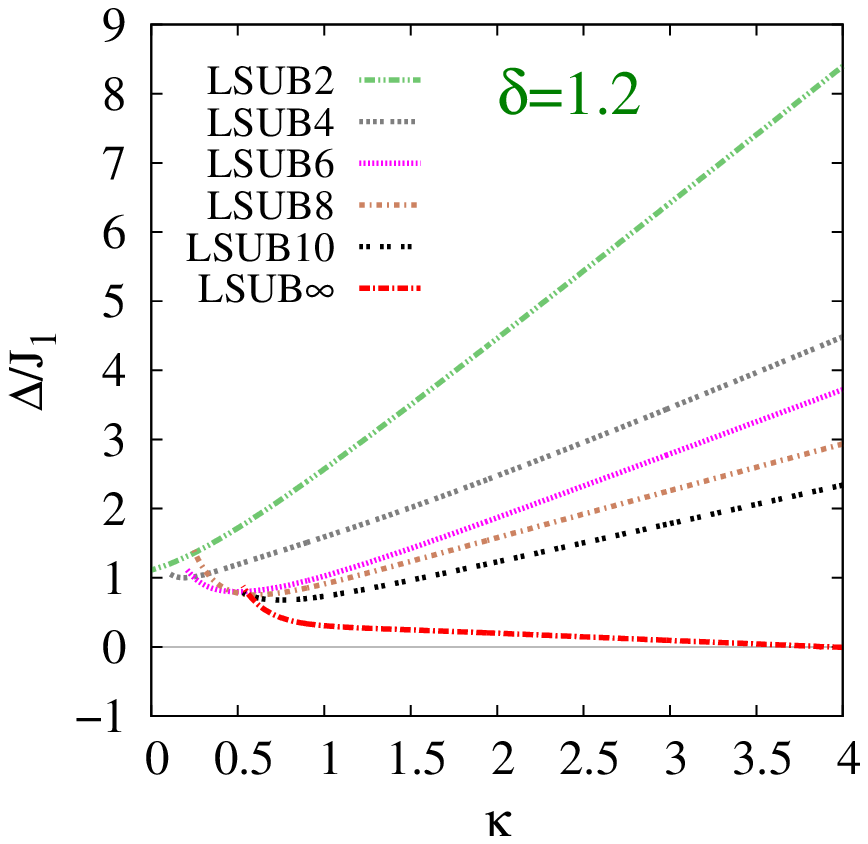}}
\hspace{-1.7cm}
\subfigure[]{\includegraphics[width=6.1cm]{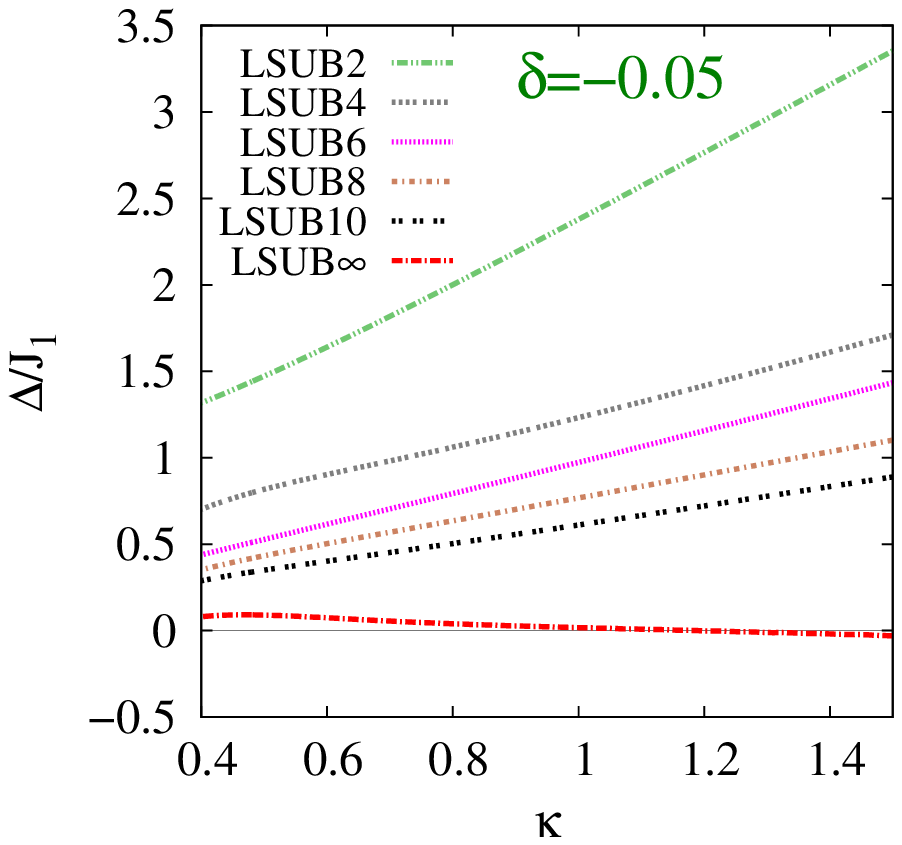}}
}
\caption{CCM results for the excitation energy $\Delta$ (in units of $J_{1}$) from the $s^{z}_{T}=0$ ground state to the lowest-lying excited state with $s^{z}_{T}=1$ versus
  the intralayer frustration parameter,
  $\kappa \equiv J_{2}/J_{1}$, for the spin-$\frac{1}{2}$
  $J_{1}$--$J_{2}$--$J_{1}^{\perp}$ model on the bilayer honeycomb
  lattice (with $J_{1}>0$), for three selected values of the scaled
  interlayer exchange coupling constant, $\delta \equiv J_{1}^{\perp}/J_{1}$:
  (a) $\delta=0$, (b) $\delta=1.2$, and (c) $\delta=-0.05$.  Results
  based on the N\'{e}el-II state on each monolayer (and the two layers coupled so that NN spins between them are antiparallel to one another, even in case (c) where $\delta<0$) as CCM model state are shown in LSUB$n$
  approximations with $n=2,4,6,8$, and $10$, (and also with
  $n=12$ for the special case $\delta=0$ of the $J_{1}$--$J_{2}$
  monolayer), together with the LSUB$\infty$ extrapolated results
  based on Eq.\ (\ref{Eq_spin_gap}) and the LSUB$n$ data sets
  $n=\{2,6,10\}$.  For the case $\delta=0$ only we also show the
  corresponding LSUB$\infty'$ extrapolation based on the LSUB$n$ data set
  $n=\{4,8,12\}$.}
\label{Egap_raw_extrapo_fix-J1perp_neel-II}
\end{center}
\end{figure*}  
The same sort of LSUB$\infty$ extrapolated data that is shown in Fig.\
\ref{M_raw_extrapo_fix-J2_neel-II} is also displayed in the composite
Fig.\ \ref{M_J2fix-selective_extrapo_neel-II} where we show $M$ versus
$\delta$ curves for a variety of values of the intralayer frustration
parameter $\kappa$.
We see very clearly that N\'{e}el-II order can exist only for values of
$\kappa$ in the range
$\kappa^{{\rm min}}_{2} < \kappa < \kappa^{{\rm max}}_{2}$.  The corresponding values of
$\delta$ at which the N\'{e}el-II LRO disappears last are also
observed to be $\delta^{l}_{2} \approx 0.2$ and
$\delta^{u}_{2} \approx 1.2$, in accord with what was discussed
previously.  Similarly, the respective values of $\kappa$ are also
seen to be $\kappa^{{\rm min}}_{2} \approx 0.35$ and
$\kappa^{{\rm max}} \approx 3.34$, again as already discussed above.

\begin{figure*}
\begin{center}
\mbox{
\hspace{-1.0cm}
\subfigure[]{\includegraphics[width=6.1cm]{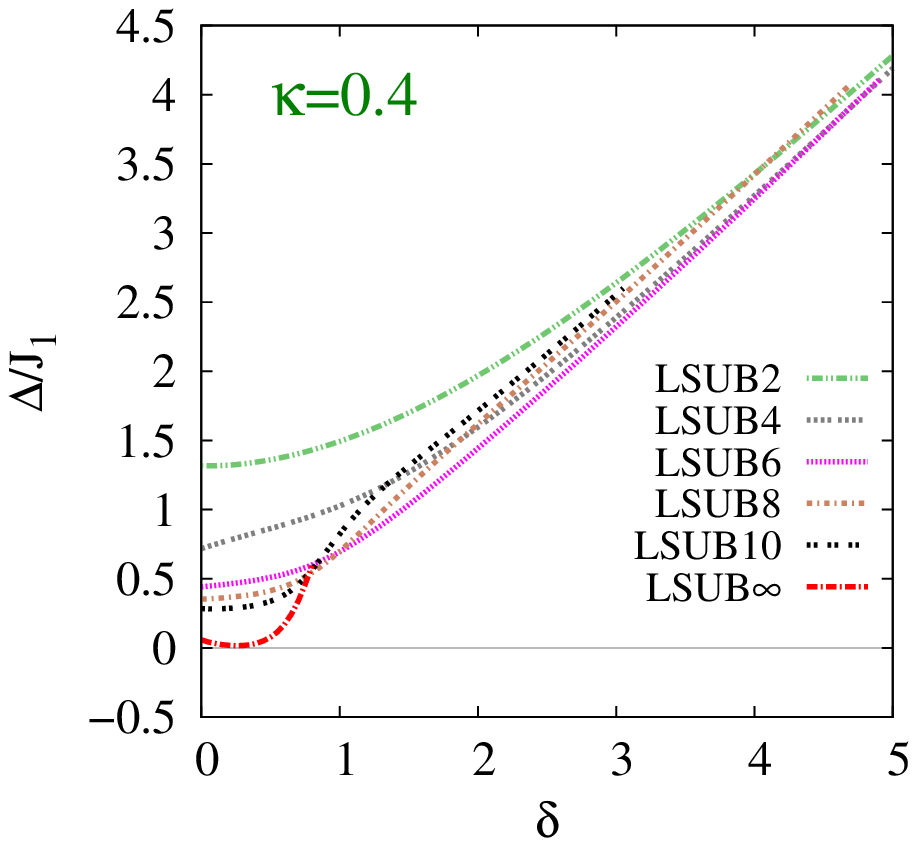}}
\hspace{-1.7cm}
\subfigure[]{\includegraphics[width=6.1cm]{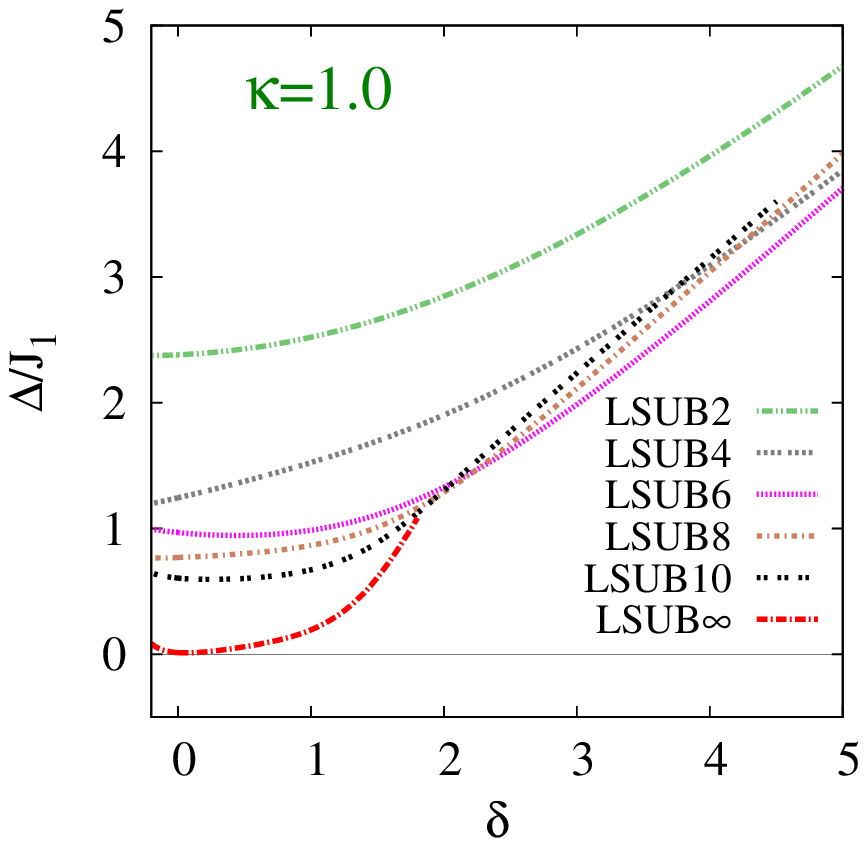}}
\hspace{-1.7cm}
\subfigure[]{\includegraphics[width=6.1cm]{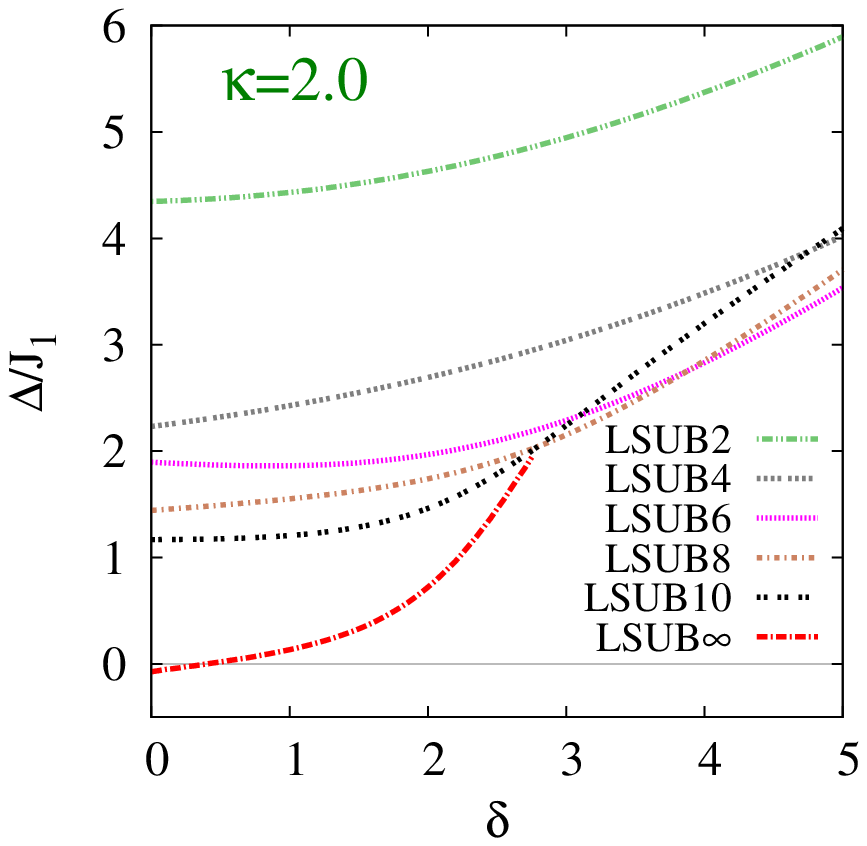}}
}
\caption{CCM results for the excitation energy $\Delta$ (in units
  of $J_{1}$) from the $s^{z}_{T}=0$ ground state to the lowest-lying excited state with $s^{z}_{T}=1$ versus the scaled interlayer exchange coupling constant,
  $\delta \equiv J_{1}^{\perp}/J_{1}$, for the spin-$\frac{1}{2}$
  $J_{1}$--$J_{2}$--$J_{1}^{\perp}$ model on the bilayer honeycomb
  lattice (with $J_{1}>0$), for three selected values of the
  intralayer frustration parameter, $\kappa \equiv J_{2}/J_{1}$: (a)
  $\kappa=0.4$, (b) $\kappa=1.0$, and (c) $\kappa=2.0$.  Results based
  on the N\'{e}el state on each monolayer (and the two layers coupled so that NN spins between them are antiparallel to one another) as CCM model state are shown in LSUB$n$
  approximations with $n=2,4,6,8$, and $10$, together with the corresponding
  LSUB$\infty$ extrapolated result based on Eq.\ (\ref{Eq_spin_gap})
  and the LSUB$n$ data sets $n=\{2,6,10\}$.  Note that, since the LSUB$n$ curves exhibit unphysical crossings deep into the ``unphysical'' region where the N\'{e}el-II model state becomes less appropriate, the LSUB$\infty$ extrapolated curves are shown over limited ranges of values of $\delta$.}
\label{Egap_raw_extrapo_fix-J2_neel-II}
\end{center}
\end{figure*}

We consider next our CCM results for the excitation energy $\Delta$ to the lowest-lying state in the $s_{T}^{z}=1$ sector.
Thus, we first show, in Fig.\
\ref{Egap_raw_extrapo_fix-J1perp_neel-II}, the corresponding results
for $\Delta$ to those shown in Fig.\
\ref{M_raw_extrapo_fix-J1perp_neel-II} for the N\'{e}el-II magnetic
order parameter $M$, for the same three fixed values of the interlayer
coupling parameter $\delta$.
Our LSUB$n$ results for the spin-$\frac{1}{2}$ honeycomb-lattice
monolayer (i.e., for the case $\delta=0$) are shown in Fig.\
\ref{Egap_raw_extrapo_fix-J1perp_neel-II}(a).  Once again, for this
limiting case we can perform LSUB$n$ approximations with $n \leq 12$,
whereas for the cases $\delta \neq 0$ we are constrained to those with
$n \leq 10$.  Just as in Fig.\
\ref{M_raw_extrapo_fix-J1perp_neel-II}(a) for $M$, so in Fig.\
\ref{Egap_raw_extrapo_fix-J1perp_neel-II}(a) for $\Delta$ we show the
two extrapolations: LSUB$\infty$ based on Eq.\ (\ref{Eq_spin_gap}) and
the input LSUB$n$ data set $n=\{2,6,10\}$, and the corresponding
LSUB$\infty'$ extrapolation based on the LSUB$n$ data set
$n=\{4,8,12\}$.  In overall terms the two extrapolations are in good
agreement.  In particular, both give results for $\Delta$ that are
zero, up to small numerical errors, over the entire range shown with
$\kappa > \kappa^{c,>}_{2}(0)$.  This is good evidence that the QCP at
$\kappa^{c,>}_{2}(0)$ is between two gapless states, compatible with
the hypothesis that, at least for the monolayer ($\delta=0$), the
transition is from one quasiclassical state, namely the N\'{e}el-II
state, to another, presumably a state with spiral order.  Conversely,
there is some slight evidence that at the QCP at
$\kappa^{c,<}_{2}(0)$, the transition might be to a gapped state,
presumably a VBC state.

In Figs.\ \ref{Egap_raw_extrapo_fix-J1perp_neel-II}(b) and
\ref{Egap_raw_extrapo_fix-J1perp_neel-II}(c) respectively we show
results for $\Delta = \Delta(\kappa)$ for the honeycomb-lattice
bilayer, with the value $\delta=1.2$ and $\delta=-0.05$ for the
interlayer coupling parameter.  Once again, in both cases it seems
that the QCP at $\kappa^{c,>}_{2}(\delta)$ is from the N\'{e}el-II
state to another gapless state.  By contrast, our results for both of
the values of $\delta$ shown in Figs.\
\ref{Egap_raw_extrapo_fix-J1perp_neel-II}(b) and
\ref{Egap_raw_extrapo_fix-J1perp_neel-II}(c) indicate that the QCP at
$\kappa^{c,<}_{2}(\delta)$ is from the quasiclassical N\'{e}el-II
state to a (nonclassical) gapped state.

\begin{figure*}[t]
\begin{center}
\mbox{
\hspace{-0.7cm}
\subfigure[]{\includegraphics[width=5.5cm]{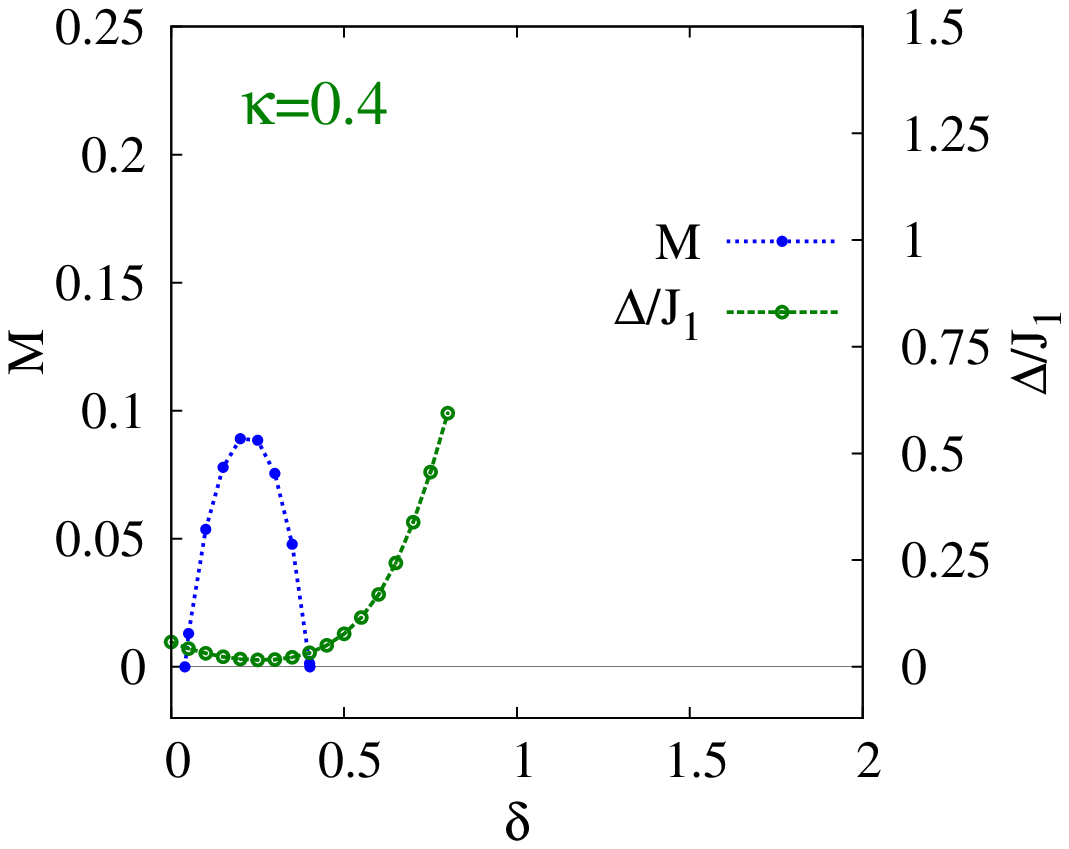}}
\hspace{-1.1cm}
\subfigure[]{\includegraphics[width=5.5cm]{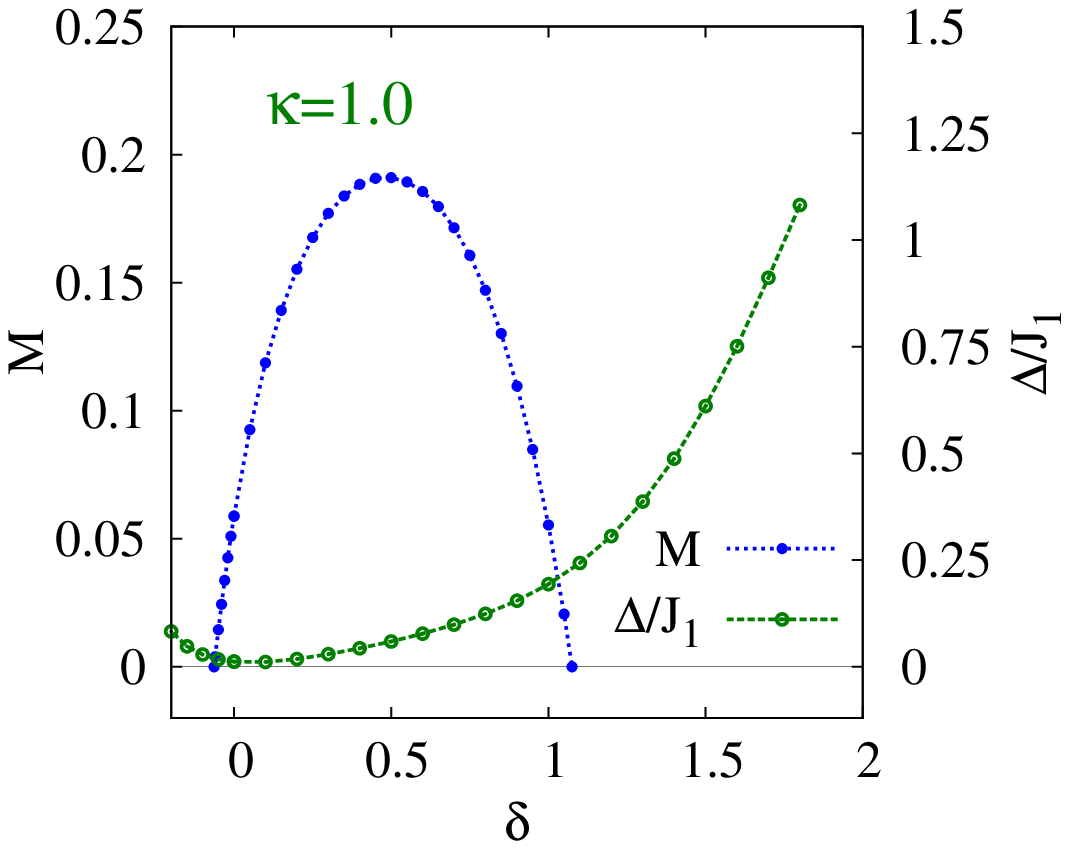}}
\hspace{-1.1cm}
\subfigure[]{\includegraphics[width=5.5cm]{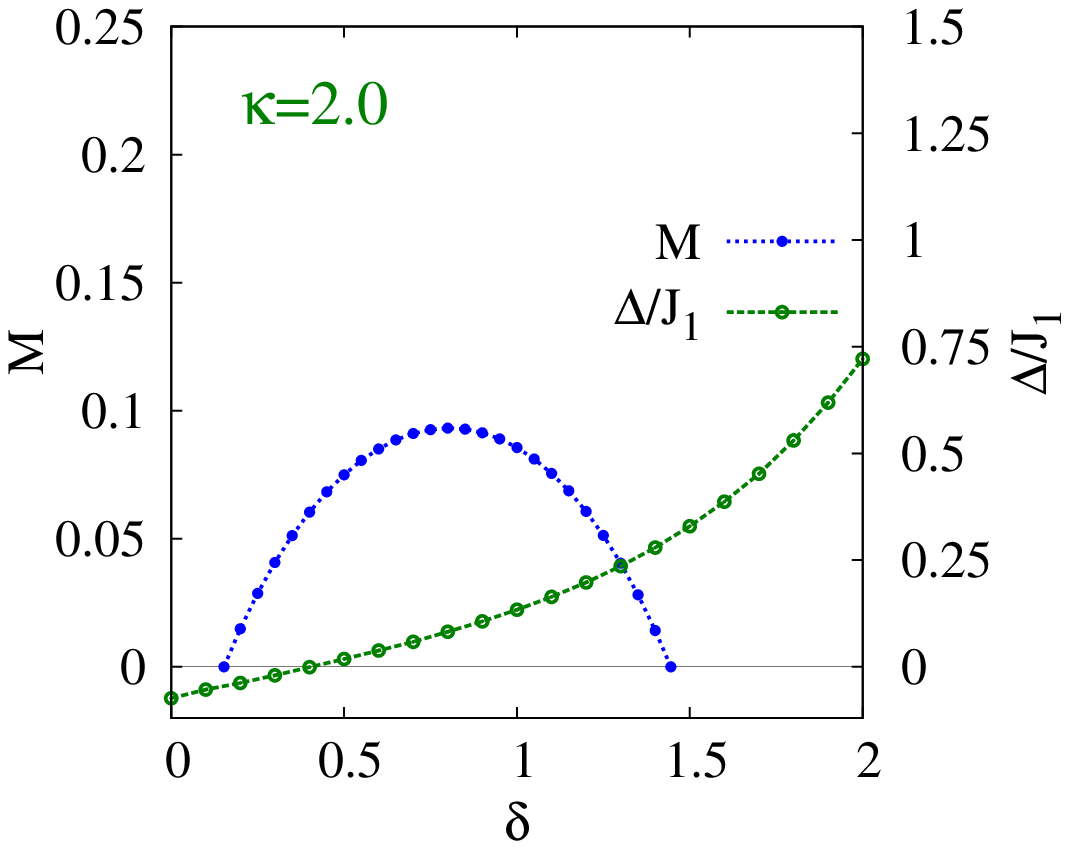}}
}
\caption{Juxtaposed CCM results for the magnetic order parameter $M$ (left
  scale) and the excitation energy $\Delta$ (in units of $J_{1}$, right
  scale)  from the $s^{z}_{T}=0$ ground state to the lowest-lying excited state with $s^{z}_{T}=1$ versus the scaled interlayer exchange coupling constant,
  $\delta \equiv J_{1}^{\perp}/J_{1}$, for the spin-$\frac{1}{2}$
  $J_{1}$--$J_{2}$--$J_{1}^{\perp}$ model on the bilayer honeycomb
  lattice (with $J_{1}>0$), for three selected values of the
  intralayer frustration parameter, $\kappa \equiv J_{2}/J_{1}$: (a)
  $\kappa=0.4$, (b) $\kappa=1.0$, and (c) $\kappa=2.0$.  Extrapolated
  results for $M$ and $\Delta$ are shown from using Eqs.\
  (\ref{M_extrapo_frustrated}) and (\ref{Eq_spin_gap}), respectively,
  with the corresponding LSUB$n$ data sets with $n=\{2,6,10\}$ in
  each case, based on the N\'{e}el-II state on each monolayer (and the two layers coupled so that NN spins between them are antiparallel to one another) as the CCM model state.}
\label{M_Egap_extrapo_fix-J2_neel-II}
\end{center}
\end{figure*}

The effect of the interlayer coupling on the excitation energy $\Delta$ to the lowest-lying state in the $s_{T}^{z}=1$ sector
is also shown in Fig.\ \ref{Egap_raw_extrapo_fix-J2_neel-II} for the
same three different values of the intralayer frustration parameter
$\kappa$ as have been shown in Fig.\
\ref{M_raw_extrapo_fix-J2_neel-II} for the N\'{e}el-II magnetic order
parameter $M$.  The LSUB$\infty$ extrapolations shown in each case are
compatible with the phase of the system being gapless over the
respective ranges of values for the interlayer coupling parameter
$\delta$ for which N\'{e}el-II magnetic LRO survives according to the
corresponding LSUB$\infty$ extrapolations for $M$ shown in Fig.\
\ref{M_raw_extrapo_fix-J2_neel-II}.  We see very clearly in each case
that at the upper critical point $\delta^{c,>}_{2}(\kappa)$ the
transition is from the gapless quasiclassical N\'{e}el-II state to a
(nonclassical) gapped state, which is presumably again a state with VBC
order.  The nature of the corresponding transitions at the lower
critical points $\delta^{c,<}_{2}(\kappa)$ can be seen more clearly
from Fig.\ \ref{M_Egap_extrapo_fix-J2_neel-II}, where we juxtapose our
extrapolated LSUB$\infty$ results for $M(\delta)$ and $\Delta(\delta)$
on the same graph, for each of the three fixed values of $\kappa$
shown in Figs.\ \ref{M_raw_extrapo_fix-J2_neel-II} and
\ref{Egap_raw_extrapo_fix-J2_neel-II}.

Of course, {\it all} quasiclassical states with magnetic LRO
spontaneously break the continuous symmetry of the Hamiltonian of the
system of Eq.\ (\ref{H_eq}) under rotations in spin space.  Hence, by
Goldstone's theorem, the N\'{e}el-II state must have soft (Goldstone)
excitation modes with a zero energy gap.  The (relatively small but)
nonzero values of $\Delta$ shown in Fig.\
\ref{M_Egap_extrapo_fix-J2_neel-II} in the respective regions where
the N\'{e}el-II magnetic order parameter $M$ is nonzero, must hence be
taken as indicative of the errors associated with extrapolating the
CCM LSUB$n$ data for the excitation energy $\Delta$.  These errors are
largest in Figs.\ \ref{M_Egap_extrapo_fix-J2_neel-II}(a) and
\ref{M_Egap_extrapo_fix-J2_neel-II}(b).  However, even in these cases,
a closer inspection of Fig.\ \ref{Egap_raw_extrapo_fix-J2_neel-II},
particularly observing the LSUB10 results for both cases, clearly
reveals that the deviations from zero values for the extrapolated excitation energies
are very likely to be due to the (unavoidable) inclusion of the LSUB2
results in the extrapolations.  Despite these numerical shortcomings,
our results point to the transition at $\delta^{c,<}_{2}(\kappa)$
being to a gapped (and, hence, presumably a VBC) state for
$\kappa=0.4$ and $1.0$, but to a gapless (and, hence, probably a
spiral quasiclassical) state for $\kappa=2.0$.

\begin{figure}[!t]
\begin{center}
  \includegraphics[width=10cm]{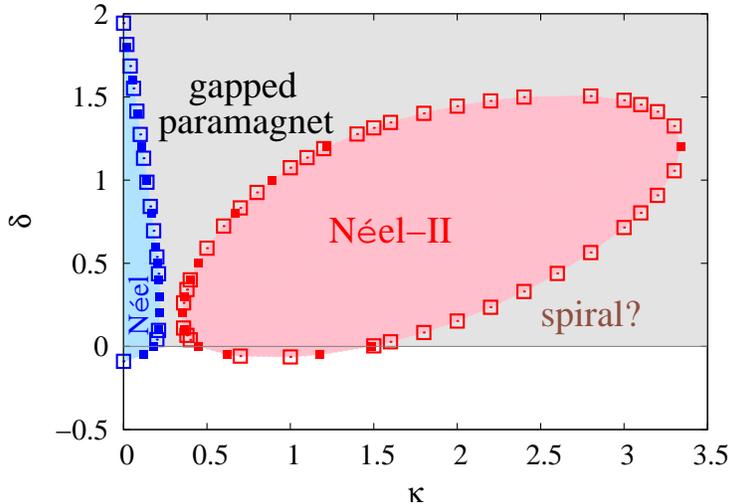}
\end{center}
  \caption{$T=0$ phase diagram of the spin-$\frac{1}{2}$
    $J_{1}$--$J_{2}$--$J_{1}^{\perp}$ model on the bilayer honeycomb
    lattice with $J_{1}>0$, $\delta \equiv J_{1}^{\perp}/J_{1}$, and
    $\kappa \equiv J_{2}/J_{1}$.  The blue and pink shaded regions are the
    quasiclassical phases with AFM N\'{e}el and N\'{e}el-II orders in each monolayer (and the two layers coupled so that NN spins across them are anti-aligned), respectively, while in the grey shaded region quasiclassical collinear order is absent.  In the white unshaded  region (where $\delta < 0$) there will be larger stable regions of N\'{e}el and N\'{e}el-II ordering on each monolayer but with the two layers coupled so that NN spins between them are parallel to one another, as well as phases with no collinear magnetic order.  Such phases have not been investigated here.  The filled and empty square
    symbols are points at which the
    extrapolated GS magnetic order parameter $M$ for the two quasiclassical AFM phases
    vanishes, for specified values of $\delta$ and $\kappa$,
    respectively.  In each
    case the N\'{e}el or N\'{e}el-II state on each monolayer (and the two layers coupled so that NN spins between them are antiparallel to one another, even when $\delta<0$) is used as CCM model state, and Eq.\
    (\ref{M_extrapo_frustrated}) is used for the extrapolations with
    the corresponding LSUB$n$ data sets $n=\{2,6,10\}$.}
\label{phase_diag_neel_neel-II}
\end{figure}

As we have noted elsewhere (and see, e.g., Ref.\
\cite{Bishop:2017_honeycomb_bilayer_J1J2J1perp}) for calculations
performed within the CCM methodology, the vanishing of the magnetic
order parameter $M$ almost always gives a considerably more accurate
estimate for the position of a QCP from a gapless to a gapped state of
a system than the opening up of a nonzero value for $\Delta$.  Thus, usually at
a QCP where the extrapolated value for $M$ vanishes, the corresponding
value for the slope of the curve for $M$ as a function of the relevant
coupling parameter is nonzero, precisely as is the case in Figs.\
\ref{M_raw_extrapo_fix-J1perp_neel-II},
\ref{M_raw_extrapo_fix-J2_neel-II}, and
\ref{M_J2fix-selective_extrapo_neel-II}.  By contrast, however, the
corresponding extrapolated curves for $\Delta$ generally depart from
being zero at the respective QCPs with zero slope, again precisely as
we see here in Figs.\ \ref{Egap_raw_extrapo_fix-J1perp_neel-II} and
\ref{Egap_raw_extrapo_fix-J2_neel-II}.  It is then inevitable that our
CCM estimates for any such QCP from results obtained for $\Delta$ have
appreciably larger associated errors than those obtained from $M$.

Thus, in Fig.\ \ref{phase_diag_neel_neel-II} we show our final results
for the $T=0$ quantum phase diagram of the model in the $\kappa\delta$
half-plane with $\kappa > 0$, using our
LSUB$\infty$ results for the points where the magnetic order parameter
vanishes to demarcate the phase boundaries of the two collinear AFM
phases, in each case with the two layers coupled so that NN spins between them are anti-aligned.
Earlier results for the N\'{e}el phase
\cite{Bishop:2017_honeycomb_bilayer_J1J2J1perp} in the region
$\delta > 0$ are also supplemented here with values $\delta < 0$,
using exactly the same CCM framework as used here for the N\'{e}el-II
phase, except that the N\'{e}el state on each monolayer is used as the
CCM model state.  Different symbols are used in Fig.\
\ref{phase_diag_neel_neel-II} to distinguish between points on the
phase boundaries that have been obtained from calculations at fixed
values of $\delta$ (such as those in Fig.\
\ref{M_raw_extrapo_fix-J1perp_neel-II} for the N\'{e}el-II state) and
those that have been obtained from respective calculations at fixed
values of $\kappa$ (such as those in Fig.\
\ref{M_raw_extrapo_fix-J2_neel-II} for the N\'{e}el-II state).  The
fact that these two sets of critical points lie so accurately on a
smooth common boundary curve for each quasiclassical collinear state is
an excellent internal check on the accuracy of the extrapolation
scheme of Eq.\ (\ref{M_extrapo_frustrated}), which has been used to
obtain them.  The results are summarized and discussed in more detail
in Sec.\ \ref{discuss_summary_sec}.

\section{Discussion and Summary}
\label{discuss_summary_sec}
The $T=0$ quantum phase diagram of the spin-$\frac{1}{2}$
$J_{1}$--$J_{2}$--$J_{1}^{\perp}$ model on an $AA$-stacked honeycomb
bilayer lattice has been investigated here within the computational
framework of the CCM.  We have focussed attention on calculating the
complete phase boundaries of the two collinear quasiclassical AFM
phases, namely those with N\'{e}el and N\'{e}el-II magnetic LRO on
each monolayer (with the two layers coupled so that NN spins between them are anti-aligned), in the half-plane $\kappa > 0$, where intralayer
frustration is present, of the complete parameter space spanned by the
intralayer frustration parameter, $\kappa \equiv J_{2}/J_{1}$, and the
interlayer coupling strength, $\delta \equiv J_{1}^{\perp}/J_{1}$, for
the case of AFM coupling $J_{1} > 0$.  The CCM has been used because
it has the distinct dual advantages of satisfying both the Goldstone
linked cluster theorem and the Hellmann-Feynman theorem at every level of
approximation that we use.  In these two important regards the method
is essentially unequalled by any other technique of {\it ab initio}
quantum many-body theory that can be applied to such spin-lattice
systems as we study here.  One consequence is that we have been able
to perform all of our calculations in the thermodynamic (infinite
lattice, $N \to \infty$) limit from the outset, thereby obviating the
need for any finite-size scaling of our results.  Since such scaling
is usually an important source of errors in competing methods, and
since finite systems often do not share the same GS ordering as their
infinite counterparts, it is a real strength of our calculations that
we have been able to circumvent these issues.

Nevertheless, of course, we have necessarily had to make
approximations.  However, we have done so within the context of a
well-defined hierarchy of truncations for the CCM multispin
correlations that are retained (viz., the so-called LSUB$n$ scheme),
which has been rigorously tested on many previous occasions in
applications to a large number of strongly correlated and highly
frustrated quantum spin-lattice models.  The approximations are
guaranteed to become exact in the limit that the truncation order
becomes infinite ($n \to \infty$), and our sole approximation for the
calculation of any physical parameter is to perform this extrapolation
on the corresponding LSUB$n$ sequences of approximants that are
computationally feasible to perform.  For the present model we have
been able to implement the method for calculations of both the
magnetic order parameter $M$ and the excitation energies $\Delta$ from the $s^{z}_{T}=0$ collinear ground states to the lowest-lying respective states in the conserving $s^{z}_{T}=1$ sectors, to very
high orders, namely those with $n \leq 10$.

While there exist by now very well-tested and much studied LSUB$n$
extrapolation schemes for the quantities $M$ and $\Delta$ among
others, an additional complication arises in the case of the honeycomb
lattice, which manifests itself as a $(4m-2)/4m$ staggering effect in
the sequences of LSUB$n$ approximants, and which we have discussed in
detail in Sec.\ \ref{ccm_sec}.  Since its origin almost certainly lies
in the non-Bravais nature of the honeycomb lattice, it is important to
realize that its effect is unavoidable and {\it must} be taken into
account when high accuracy is required.  Indeed, it is a testament to
the power and accuracy of the CCM that this additional staggering has
been clearly observed here.  Naturally, it is possible, even likely, that the same
or related staggering effects also occur in other calculational
schemes, where they have perhaps been overlooked hitherto.

For reasons that we have enumerated, our most accurate calculations
for the phase boundaries on which N\'{e}el or N\'{e}el-II magnetic LRO
melts come from the points where the respective order parameter $M$
vanishes.  Nevertheless, our results for $\Delta$ provide excellent
independent corroboration of the regions in which quasiclassical
magnetic order is present (i.e., where $\Delta$ vanishes), while also
giving some additional information on whether the transition from one
of the collinear AFM states at a given point on a boundary is to a
gapped or a gapless state.

It is perhaps worth noting at this point that we have not attempted to
implement any wave vector optimization to search for a {\it lowest} spin
gap.  In the first place any such attempt would add very significantly
to the computational burden of what is already a set of calculations
that push at the limit of what can be performed at the very high
LSUB10 level of approximation that is required for accuracy in these
calculations, even with the use of powerful supercomputing resources.
Indeed, any such wave vector optimization, at the required level of
implementation to give accurate, meaningful CCM results for this
model, is definitely beyond computational reach at this point.
Secondly, our need is certainly less ambitious, since our ES results are only used here both to corroborate the intrinsically
more accurate GS order parameter results, and to try to shed some more
light on the nature of the states proximate to the quasiclassical AFM
states under consideration.

We note again that the stability of the N\'{e}el-II phase that we have
observed over a large parameter regime in the $\kappa\delta$ plane
cannot be explained on classical grounds.  Rather, for the case of a
single layer (i.e., when $\delta=0$), the N\'{e}el-II state exists for
a certain range of values of $\kappa$ for the spin-$\frac{1}{2}$
$J_{1}$--$J_{2}$ model on the hexagonal lattice in which the classical
model has spiral order.  However, as is generically always the case,
quantum fluctuations tend to favor collinear order over non-collinear
order.  Hence, for the spin-$\frac{1}{2}$ $J_{1}$--$J_{2}$ model, the
collinear N\'{e}el-II state, which lies close in energy classically to
the spiral ground state over a regime around $\kappa = \frac{1}{2}$,
and which is actually degenerate in energy with the spiral states
exactly at the classical critical point at $\kappa=\frac{1}{2}$, then
becomes promoted by quantum fluctuation to be the lowest-energy state
in a range of values around this point.  This (rather fragile)
stability of the AFM N\'{e}el-II phase is then initially enhanced,
exactly as for the simpler AFM N\'{e}el state, by turning on an
interlayer coupling, $\delta > 0$, in the spin-$\frac{1}{2}$
$J_{1}$--$J_{2}$--$J_{1}^{\perp}$ model on the bilayer honeycomb
lattice, when the ordering increases in both cases.  The reason is
presumably the same for both phases, and must again be a quantum
effect, since at the classical level such an interlayer coupling plays
no role whatsoever.  Thus, for the spin-$\frac{1}{2}$ model a small
interlayer coupling clearly enhances the AFM LRO of both
quasiclassical (N\'{e}el and N\'{e}el-II) phases, as can clearly be
seen from (Fig.\ \ref{M_J2fix-selective_extrapo_neel-II} and) Fig.\
\ref{phase_diag_neel_neel-II}.  However, as $\delta$ is increased
further, any quasiclassical magnetic ordering in each layer now starts
to compete with an IDVBC state formed as a product of
interlayer singlet dimers between NN pairs on the $AA$-stacked
bilayer.  This increasing competition then leads ultimately, for both
the N\'{e}el and N\'{e}el-II phases, to a weakening of their
respective magnetic ordering, leading to the reentrant behavior seen
clearly for both phases in Fig.\ \ref{phase_diag_neel_neel-II}.  The
reentrant effect is rather similar in both phases, with a maximum
enhancement of the AFM magnetic LRO at a value $\delta \approx 0.2$,
for the N\'{e}el case at an upper critical value
$\kappa_{1}^{{\rm max}} \approx 0.215$ and for the N\'{e}el-II case at
a lower critical value $\kappa_{2}^{{\rm min}} \approx 0.35$.

From our final results in Fig.\ \ref{phase_diag_neel_neel-II} it is
now easy to understand why accurate estimates for the positions of
each of the two QCPs at $\kappa^{2,<}_{c}(0)$ and
$\kappa^{2,>}_{c}(0)$, which delimit the range of values for the
frustration parameter $\kappa$ over which N\'{e}el-II magnetic LRO
exists for the honeycomb-lattice monolayer, are so difficult to
obtain.  Thus, since the $\delta=0$ axis is so close to the lower
boundary $\delta=\delta^{2,<}_{c}(\kappa)$ of N\'{e}el-II stability in each monolayer, with the two layers coupled so that NN spins between them are antiparallel to one another,
the inclusion of even a very small interlayer coupling is bound to
have a much larger effect on the corresponding estimates for the QCPs.
To a somewhat lesser extent, the same situation is also seen to be
responsible for the sensitivity in estimating the (upper) critical
point $\kappa^{1,>}_{c}(0)$ at which N\'{e}el order melts in the
honeycomb-lattice monolayer.

While it is far beyond the scope of the present investigation to enquire in detail about the
nature of the GS phases for the spin-$\frac{1}{2}$
$J_{1}$--$J_{2}$--$J_{1}^{\perp}$ model on the $AA$-stacked bilayer
honeycomb lattice outside the regions shown in Fig.\
\ref{phase_diag_neel_neel-II} where we have calculated that the model
exhibits quasiclassical collinear AFM ordering of either the N\'{e}el
or N\'{e}el-II type, we conclude with a few remarks on this issue.
Firstly, our results for the ES parameter $\Delta$, such as those
shown in Figs.\ \ref{Egap_raw_extrapo_fix-J1perp_neel-II} and
\ref{Egap_raw_extrapo_fix-J2_neel-II}, clearly indicate the presence
of a gapped paramagnetic state in part of the $\kappa \delta$
parameter space shown in Fig.\ \ref{phase_diag_neel_neel-II}.  Broadly
speaking, as indicated on the phase diagram, this gapped state exists
over all (or most) of the region between the N\'{e}el and N\'{e}el-II
islands of stability, as well as the region
$\delta>\delta^{2,>}_{c}(\kappa)$ immediately above the N\'{e}el-II island of
stability.  There is also weak evidence from results such as those
shown in Figs.\ \ref{M_Egap_extrapo_fix-J2_neel-II}(a) and
\ref{M_Egap_extrapo_fix-J2_neel-II}(b) that this paramagnetic region
also extends somewhat below the regions of stability of both the
N\'{e}el and N\'{e}el-II phases, at least for values
$\kappa \lesssim 1$.  By contrast, results from calculations such as
those shown in Figs.\ \ref{Egap_raw_extrapo_fix-J1perp_neel-II} and
\ref{M_Egap_extrapo_fix-J2_neel-II}(c) indicate that at least part of
the region below and to the right of the N\'{e}el-II island of
stability is gapless.  For reasons discussed more fully in Sec.\
\ref{model_sec}, the gapped paramagnetic region is most likely to
comprise VBC phases of different sorts, including those of the
plaquette (PVBC) and staggered dimer (SDVBC) type on each monolayer
and the interlayer dimer (IDVBC) type between the two layers.
Similarly, the gapless state in the region indicated above is likely
to be a quasiclassical state with spiral ordering.  We hope that our
preliminary findings concerning the possible regions of stability of
the paramagnetic and spiral phases for the bilayer might inspire other
calculations to investigate in more detail their regions of
stability.

\section*{ACKNOWLEDGMENT}
We thank the University of Minnesota Supercomputing Institute for the
grant of supercomputing facilities, on which the work reported here
was performed.

\section*{References}



\bibliographystyle{elsarticle-num} 
\bibliography{bib_general}





\end{document}